\documentclass[useAMS,usenatbib]{mn2e}
\usepackage{epsfig}
\usepackage{graphicx,rotating,amssymb,supertabular,longtable,lscape}
\usepackage{amsmath}
\usepackage{textcomp}

\def\mum{\textmu m}
\def\mums{\textmu m }

\title[Properties of dusty tori in SWIRE/SDSS Quasars]
{Properties of dusty tori in AGN: I. The Case of SWIRE/SDSS Quasars}

\author[Hatziminaoglou et al.]{E. Hatziminaoglou$^{1,2}$, J. Fritz$^{3}$, A. Franceschini$^{3}$,
A. Afonso-Luis$^{2}$,
\newauthor A. Hern\'{a}n-Caballero$^{2}$, I. P\'{e}rez-Fournon$^{2}$, S. Serjeant$^{4}$,
C. Lonsdale$^{5,6}$, 
\newauthor S. Oliver$^{7}$, M. Rowan-Robinson$^{8}$, D. Shupe$^{5}$, H.E. Smith$^{9}$, J. Surace$^{5}$\\
$^{1}$European Southern Observatory, Karl-Schwarzschild-Str. 2, 85748 Garching bei M\"unchen, Germany\\
$^{2}$Instituto de Astrof\'isica de Canarias, C/ V\'ia L\'actea s/n, E-38200 La Laguna, Spain\\
$^{3}$Dipartimento di Astronomia, Universita di Padova, Vicolo dell'Osservatorio 5, 35122 Padua, Italy\\
$^{4}$Centre for Astrophysics and Planetary Science, School of Physical Sciences,
University of Kent, Canterbury, Kent CT2 7NR, UK\\
$^{5}$Infrared Processing and Analysis Center, California Institute of Technology, Pasadena, CA 91125, USA\\
$^{6}$Department of Astronomy, University of Virginia, Charlottesville, VA 22904, USA\\
$^{7}$Astronomy Centre, Department of Physics and Astronomy, University of Sussex, Falmer,
Brighton BN1 9QJ, UK\\
$^{8}$Astrophysics Group, Blackett Laboratory, Imperial College London, London SW7 2BW, UK\\
$^{9}$Center for Astrophysics and Space Sciences, University of California, San Diego, La Jolla, CA 92093-0424
, USA}

\begin{document}

\maketitle

\begin{abstract}
We derive the properties of dusty tori in Active Galactic Nuclei 
(AGN) from the comparison of observed Spectral 
Energy Distributions (SEDs) of SDSS quasars and a precomputed grid of torus 
models.
The observed SEDs comprise SDSS photometry, 2MASS $J, H$, and $K$ data, whenever available
and mid-Infrared (MIR) data from the Spitzer Wide-area InfraRed Extragalactic 
(SWIRE) Survey. The adopted model is that of \cite{fritz06}. The fit is performed by 
standard $\chi^2$ minimisation, the model however can be multi-component 
comprising a stellar and a starburst components, whenever necessary.
Models with low equatorial optical depth, $\rm \tau_{9.7}$, were
allowed as well as ``traditional'' models with $\rm \tau_{9.7} \ge 1.0$, corresponding to
A$_{\rm V} \ge$ 22 and the results were compared. Fits using high optical depth tori 
models only produced dust more compactly distributed than in the configuration where all
$\rm \tau_{9.7}$ models were permitted.
Tori with decreasing dust density with the distance from the centre were favoured
while there was no clear preference for models with or without angular variation of
the dust density. 
The computed outer radii of the tori are of some tens of parsecs large but can reach, 
in a few cases, a few hundreds of parsecs.
The mass of dust, $\rm M_{Dust}$, and infrared luminosity, $\rm L_{IR}$, integrated in
the wavelength range between 1 and 1000 \mum, do not show significant variations with
redshift, once the observational biases are taken into account.
Objects with 70 \mums detections, representing 25\% of the sample, are studied separately 
and the starburst contribution (whenever present) to the IR luminosity can reach,
in the most extreme but very few cases, 80\%.
\end{abstract}


\section{INTRODUCTION}
\label{intro}

Efforts in understanding the physics at work in active galactic nuclei (AGN) 
started four decades ago.
The origin of the infrared (IR) continuum of AGN was initially a matter 
of controversy, as it could be non-thermal but could equally be due to
thermal emission from dust grains. It was long ago suggested \citep{rees69}
that IR emission radiation from Seyfert galaxies in the 2.2 -- 22 \mums wavelength
range was produced by dust grains heated by ultraviolet (UV) and
optical emission from the nucleus. 
Work carried out later on suggests the IR emission to be the reprocessed 
emission of the UV/optical radiation from the accretion disk by the particles 
composing the torus, namely silicate and graphite grains (e.g. \citealt{pier92}; 
\citealt{granato94}; 
\citealt{efstathiou95}; \citealt{nenkova02}). Various configurations of the dust distribution
geometry and compositions have been since suggested (e.g. \citealt {pier92};
\citealt{vanbemmel03}; \citealt{dullemond05}; \citealt{fritz06}; \citealt{elitzur06}).

Recent observations \citep{jaffe04} indicate that this structure might be smaller
than originally thought, but resolved in at least some nearby
systems. Interferometric measurements at 8 -- 13.5 \mums centred on
NGC~1068, a prototype Seyfert 2 galaxy,
revealed a structure with an extent of $\sim 3.4$ pc that can
be identified with the torus \citep{jaffe04}.

Dust remains essential to understanding the Seyfert 1/2 dichotomy.
The dust sublimation radius defines the outer boundary of
the broad-line region (BLR). Recent results from reverberation mapping
place the IR-emitting medium just beyond the BLR
\citep{suganuma06}. The IR emitter
presumably corresponds to the torus, that would make a type-1 object look like a
Seyfert 2 nucleus when the line of sight to the BLR is obscured by the
dusty medium. However, if the obscuring torus is ``clumpy'' rather than 
homogeneous, as suggested by \cite{krolik88} and modeled by e.g. \cite{nenkova02},
obscuration and classification of AGN becomes probabilistic.

In this paper we address the question of the distribution
of properties of dust tori around AGN, studying SDSS quasars in the fields
covered by the Spitzer Wide-area InfraRed Extragalactic
Survey (SWIRE\footnote{http://swire.ipac.caltech.edu}; 
\citealt{lonsdale03}).
We assume that tori are smooth (as opposed to ``clumpy'') 
distributions of dust, composed of graphite and silicate
grains and will limit ourselves to the use of photometry only,
spanning a large wavelength range, from $\sim$ 0.35 to 160 \mum.
Our aim is to classify the properties of the objects
under study fitting their observed Spectral Energy Distributions (SEDs), 
built by photometric
points collected from various surveys, to various model components,
including stellar templates, AGN tori and starburst emission, whenever
applicable. The quasar sample consists of all spectroscopically 
confirmed type-1 quasars in the common regions between SWIRE and SDSS 
Data Release 4 (DR4; \citealt{adelman06}).
The purpose of this work is to 
constrain the model parameters and to quantify the IR
properties of bright quasars,
within the limitations imposed by the model assumptions, the resolution
provided by the broad band photometry, and the degeneracies
resulting from the fitting procedure.

The paper is structured as follows. Section \ref{data}
describes in detail the samples. A brief description of the 
torus models and the stellar and starburst components is given in Section 
\ref{model}. Section \ref{sedfit} describes the SED fitting mechanism and the
physical parameters thus derived. The results are 
presented in Section \ref{results}, with a subsection dedicated 
to the problems of degeneracy and aliasing and their implications. 
Finally, Section \ref{discuss} presents a discussion of the results of this 
work.

\section{THE SAMPLE}
\label{data}

For the purposes of this study consistent IR data are essential, covering as
large a wavelength range as possible. SWIRE provided the astronomical 
community with unprecedented quality MIR photometric data
for over 2 million objects, including hundreds of thousands of AGN of
all types. We therefore selected the samples from regions with overlapping
SDSS DR4 and SWIRE IRAC or MIPS 24 \mums data; SWIRE ELAIS N1 and N2 and the Lockman field.
The sample comprises 278 spectroscopically confirmed SDSS 
quasars within these fields with redshifts spanning $0.06 < z < 5.2$.

The SED of each object was constructed using the
SDSS photometry ($u, g, r, i, z$), 2MASS photometry ($J, H, K$) and IR photometry
from SWIRE, i.e. IRAC 1-4 channel fluxes, MIPS 24, 70 and 160 \mums fluxes,
whenever available. 2MASS photometry from the 2MASS $\times$6 \citep{beichman03} was used,
for objects in the Lockman hole. 
The SWIRE catalogues we use throughout this work were processed by the SWIRE 
collaboration. Details about the data can be found in \cite{lonsdale04}, 
\cite{surace04} and \cite{shupe07}.
The selection of ``reliable'' sources in 70 and 160 \mums is done as follows.
All sources with detections above 5$\sigma$ are considered reliable.
The 70 and 160 \mums flux limits adopted correspond to 90\% completeness and 
generally coincide with the 5$\sigma$ noise of the images (measured from the sky rms).
They slightly vary with the field, and for SWIRE ELAIS N1 and N2 and the Lockman fields they
are of 17, 17.5 and 18 mJy for the 70 micron, and 104, 124 and 108 mJy for the 160 \mums.
Also sources with at least 3$\sigma$ detections and a 24 micron counterpart brighter than
300${\rm \mu}$Jy, within 6" and 12", for the 70 and 160 micron sources, are taken to be real.
There will be, therefore, cases where the 70 and 160 \mums fluxes will lie below the
nominal flux limits. For more details, see Vaccari et al. (in prep).

\cite{richards06} presented an extensive multiwavelength analysis of a sample of
259 objects, most of which belong to our quasar sample, also. For the colour properties 
and global SEDs, we therefore refer to this work. For a detailed
analysis of the properties of the quasars and composite SEDs in SWIRE EN1 field, see also
\cite{hatziminaoglou05}.

Tables \ref{tab:ugriz} and \ref{tab:jhkspitzer} list the coordinates, redshifts, optical, 
near- and mid-IR photometry of the 278 quasars. Here only the first 20 entries are presented,
the full tables are available as on-line material. The SWIRE fluxes missing from Table 
\ref{tab:jhkspitzer} do not indicate drop-outs but objects lying at the edges of the fields
and escaped detection in some of the SWIRE bands due to slight variations in the rastering.
As only a relatively small fraction of the
sources have been detected in 70 \mum, the 70 and 160 \mums photometry will be
presented in a separate table in Section \ref{sec:mips}.

\begin{table*}
\caption{Coordinates, redshifts and SDSS $ugriz$ photometry for the first 20 quasars, ordered by RA.}
\small
\begin{tabular}{r c c c c c c c c}
\hline
\hline
SqNr & RA & Dec & z & $u$ & $g$ & $r$ & $i$ & $z$ \\
\hline
\hline
  1   & 10:30:39.62 & +58:06:11.6 & 0.504 & 21.740$\pm$0.17 & 20.392$\pm$0.04 & 19.123$\pm$0.02 & 18.158$\pm$0.01 & 17.634$\pm$0.02 \\
  2   & 10:30:58.68 & +58:20:34.3 & 0.714 & 19.179$\pm$0.02 & 18.992$\pm$0.02 & 19.066$\pm$0.01 & 19.121$\pm$0.03 & 18.967$\pm$0.06 \\
  3   & 10:31:13.74 & +58:21:18.9 & 1.922 & 19.956$\pm$0.03 & 19.955$\pm$0.02 & 19.931$\pm$0.02 & 19.723$\pm$0.03 & 19.657$\pm$0.10 \\
  4   & 10:31:20.11 & +58:18:51.1 & 0.493 & 19.438$\pm$0.03 & 19.136$\pm$0.02 & 19.166$\pm$0.01 & 18.892$\pm$0.02 & 18.640$\pm$0.03 \\
  5   & 10:31:47.64 & +57:58:58.1 & 2.879 & 20.719$\pm$0.07 & 19.418$\pm$0.02 & 19.336$\pm$0.02 & 19.267$\pm$0.02 & 19.122$\pm$0.04 \\
  6   & 10:31:57.06 & +58:16:09.9 & 0.591 & 20.113$\pm$0.04 & 19.735$\pm$0.04 & 19.746$\pm$0.02 & 19.408$\pm$0.02 & 19.367$\pm$0.05 \\
  7   & 10:32:22.85 & +57:55:51.2 & 1.243 & 20.061$\pm$0.04 & 20.187$\pm$0.03 & 19.961$\pm$0.03 & 20.044$\pm$0.03 & 19.997$\pm$0.08 \\
  8   & 10:32:27.93 & +57:38:22.6 & 1.969 & 20.377$\pm$0.05 & 20.416$\pm$0.03 & 20.577$\pm$0.03 & 20.371$\pm$0.04 & 20.274$\pm$0.10 \\
  9   & 10:32:36.21 & +58:00:33.9 & 0.687 & 20.219$\pm$0.04 & 19.854$\pm$0.02 & 19.827$\pm$0.02 & 19.637$\pm$0.02 & 19.786$\pm$0.07 \\
  10  & 10:32:53.03 & +58:27:07.9 & 1.603 & 19.867$\pm$0.04 & 19.787$\pm$0.02 & 19.760$\pm$0.02 & 19.569$\pm$0.03 & 19.527$\pm$0.09 \\
  11  & 10:33:01.52 & +58:37:49.9 & 1.344 & 19.227$\pm$0.02 & 19.177$\pm$0.01 & 18.963$\pm$0.01 & 19.131$\pm$0.02 & 19.423$\pm$0.08 \\
  12  & 10:33:20.31 & +58:42:25.0 & 0.873 & 18.943$\pm$0.02 & 18.897$\pm$0.02 & 18.781$\pm$0.01 & 18.790$\pm$0.02 & 18.625$\pm$0.04 \\
  13  & 10:33:33.93 & +58:28:18.8 & 0.574 & 20.803$\pm$0.07 & 20.306$\pm$0.02 & 20.419$\pm$0.03 & 20.218$\pm$0.04 & 20.046$\pm$0.14 \\
  14  & 10:33:52.75 & +58:13:40.8 & 2.123 & 19.927$\pm$0.05 & 19.863$\pm$0.04 & 19.907$\pm$0.02 & 19.799$\pm$0.02 & 19.436$\pm$0.06 \\
  15  & 10:33:59.97 & +58:34:57.8 & 3.114 & 22.366$\pm$0.36 & 19.835$\pm$0.02 & 19.577$\pm$0.02 & 19.486$\pm$0.02 & 19.499$\pm$0.08 \\
  16  & 10:34:13.89 & +58:52:52.8 & 0.745 & 18.092$\pm$0.01 & 17.807$\pm$0.01 & 17.776$\pm$0.01 & 17.825$\pm$0.02 & 17.690$\pm$0.02 \\
  17  & 10:34:21.24 & +58:06:53.0 & 0.249 & 19.576$\pm$0.07 & 18.965$\pm$0.01 & 18.424$\pm$0.01 & 18.106$\pm$0.01 & 17.968$\pm$0.05 \\
  18  & 10:34:25.70 & +58:09:54.0 & 3.279 & 23.844$\pm$0.94 & 20.241$\pm$0.04 & 19.936$\pm$0.02 & 19.831$\pm$0.02 & 19.932$\pm$0.08 \\
  19  & 10:35:08.02 & +57:12:55.8 & 2.587 & 19.947$\pm$0.04 & 19.666$\pm$0.01 & 19.637$\pm$0.02 & 19.639$\pm$0.03 & 19.568$\pm$0.06 \\
  20  & 10:35:17.52 & +59:03:09.4 & 1.303 & 19.527$\pm$0.03 & 19.656$\pm$0.02 & 19.280$\pm$0.02 & 19.316$\pm$0.02 & 19.519$\pm$0.09 \\
\hline
\hline
\end{tabular}
\label{tab:ugriz}
\end{table*}

\begin{table*}
\caption{2MASS photometry, IRAC (in ${\rm \mu}$Jy) and MIPS 24 \mums fluxes (in mJy)
for the first 20 quasars, ordered by RA. }
\small
\begin{tabular}{r c c c c c c c c}
\hline
\hline
SqNr & $J$ & $H$ & $K$ & S$_{3.6}$ [${\rm \mu}$Jy] & S$_{4.5}$ [${\rm \mu}$Jy] & S$_{5.8}$ [${\rm \mu}$Jy] & S$_{8.0}$ [${\rm \mu}$Jy] & S$_{24}$ [mJy]\\
\hline
\hline
  1 & 15.990$\pm$0.02 & 14.945$\pm$0.02 & 13.773$\pm$0.01 & 5348.3$\pm$6.89 & 6867.3$\pm$5.97 &9017.2$\pm$20.88 &11468.0$\pm$11.24&	    --\\
  2 &       --	&	--	  &	  --	    & 390.04$\pm$2.74 & 501.55$\pm$3.03 &	--	  & 981.08$\pm$9.24 &	    --\\
  3 &       --	&	--	  &	  --	    & 71.31$\pm$0.84  & 106.62$\pm$1.24 & 200.51$\pm$4.37 & 306.49$\pm$5.78 &	    --\\
  4 & 17.501$\pm$0.07 & 16.529$\pm$0.09 & 15.776$\pm$0.07 & 589.85$\pm$2.39 & 688.12$\pm$2.18 & 864.29$\pm$7.74 & 985.89$\pm$5.82 &	    --\\
  5 &       --	&	--	  &	  --	    & 100.28$\pm$1.07 & 115.14$\pm$1.15 & 167.05$\pm$5.19 & 298.72$\pm$5.19 & 1.019$\pm$0.01\\
  6 &       --	&	--	  &	  --	    & 227.3$\pm$1.24  & 290.32$\pm$1.38 & 414.32$\pm$5.1  & 565.18$\pm$4.97 & 2.534$\pm$0.03\\
  7 &       --	&	--	  &	  --	    & 116.49$\pm$0.93 & 158.44$\pm$1.42 & 204.23$\pm$4.41 & 284.76$\pm$5.78 & 0.612$\pm$0.01\\
  8 &       --	&	--	  &	  --	    & 58.28$\pm$1.66  &       --	& 162.75$\pm$8.58 &	  --	    & 0.490$\pm$0.01\\
  9 &       --	&	--	  &	  --	    & 171.56$\pm$1.0  & 245.13$\pm$1.36 & 345.04$\pm$4.61 & 475.43$\pm$4.95 & 1.165$\pm$0.01\\
 10 &       --	&	--	  &	  --	    & 95.53$\pm$1.09  & 142.35$\pm$1.35 & 198.96$\pm$5.38 & 359.6$\pm$5.91  & 0.877$\pm$0.03\\
 11 &       --	&	--	  &	  --	    & 273.12$\pm$2.26 & 419.03$\pm$2.97 & 592.55$\pm$8.17 & 857.97$\pm$7.6  &	    --\\
 12 & 17.505$\pm$0.09 & 17.070$\pm$0.16 & 16.437$\pm$0.16 & 548.98$\pm$4.46 & 590.21$\pm$3.36 &922.55$\pm$15.01 & 1163.0$\pm$7.53 &	    --\\
 13 &       --	&	--	  &	  --	    & 168.42$\pm$1.5  & 187.89$\pm$2.07 & 221.55$\pm$5.31 & 239.63$\pm$5.96 & 0.741$\pm$0.01\\
 14 &       --	&	--	  &	  --	    & 73.37$\pm$0.78  & 91.090$\pm$0.97 & 151.93$\pm$3.48 & 224.17$\pm$4.65 & 0.576$\pm$0.01\\
 15 &       --	&	--	  &	  --	    & 65.22$\pm$1.17  & 84.420$\pm$1.25 & 148.57$\pm$6.01 & 286.36$\pm$5.0  & 1.050$\pm$0.01\\
 16 & 16.603$\pm$0.04 & 16.132$\pm$0.08 & 15.374$\pm$0.07 &1331.7$\pm$6.95  & 1987.1$\pm$7.65 &2734.3$\pm$24.62 &4106.1$\pm$16.07 &	      --\\
 17 & 17.114$\pm$0.07 & 16.323$\pm$0.09 & 15.688$\pm$0.09 & 356.5$\pm$2.23  & 422.01$\pm$2.56 & 458.92$\pm$7.61 & 1147.9$\pm$6.85 & 3.309$\pm$0.01\\
 18 &       --  &       --	  &	  --	    & 51.74$\pm$0.71  & 60.800$\pm$1.08 & 72.440$\pm$3.95 & 128.3$\pm$5.58  & 0.511$\pm$0.01\\
 19 &       --	&	--	  &	  --	    &	    --        &       --	&	--	  &	  --	    & 0.621$\pm$0.01\\
 20 &       --	&	--	  &	  --	    & 204.23$\pm$2.79 & 326.77$\pm$3.73 & 412.04$\pm$11.06& 634.96$\pm$9.99 &	    --\\
\hline
\end{tabular}
\label{tab:jhkspitzer}
\end{table*}

\section{THE MODEL COMPONENTS}
\label{model}
The observed SED of a galaxy, from the UV to the far-IR (FIR), can in principle be built as the
sum of three distinct components: stars, which emit most of their
power in the optical, hot dust, heated by accretion onto a supermassive black hole, whose emission
peaks in the MIR (from a few to some tens of microns) and cold dust, mainly heated by
star formation.

\subsection{The torus component}
\label{sec:tori}

The main focus of this work is the dusty torus,
for which we are seeking to determine the properties for the individual
objects and for the sample as a whole. For the purposes of this work we assume 
a continuous dust distribution around
the central sources (an accreting black hole), 
consisting of silicate and graphite grains, confined in a toroidal shape, 
as opposed to ``clumpy'' tori models (see e.g. \citealt{nenkova02}). For a detailed description of 
the model we refer to \cite{fritz06}. Here, however, we summarize some of the 
major characteristics of the model.

The central source is assumed to be point-like and its emission isotropic. Its 
spectral energy distribution is defined by means of a composition of power laws 
(i.e. $\lambda L(\lambda) \propto \lambda^i$) with different values for the spectral 
index $i$. We adopted values found in the literature (e.g. \citealt{granato94} or 
\citealt{schartmann05}), 
namely: $i=1.2$ for $0.001 < \lambda < 0.03$,  $i=0$ for $0.03 < \lambda < 0.125$,
$i=-0.5$ and $i=-1.0$ for $0.125 < \lambda < 10$, and a Rayleigh-Jeans decline ($i=-3.0$) is
assumed longward of $\lambda =10$, where $\lambda$ is given in \mums.

The torus geometry adopted to describe the shape and the spatial distribution of 
dust is the so-called ``flared disk'' (see e.g. \citealt{efstathiou95}, 
\citealt{manske98} and \citealt{vanbemmel03}), that is a sphere with the 
polar cones removed. Its size is defined by its outer radius, R$_{out}$, and the 
opening angle, $\Theta$, of the torus itself.
The dust components that dominate both the absorption and the emission of radiation
are graphite and silicate. The location of the inner radius, 
R$_{in}$, depends both on the sublimation temperature of the dust grains (1500
and 1000 K, for graphite \citep{barvainis87} and silicate \citep{granato94}, 
respectively) and on the strength of the accretion luminosity. 
We adopted the absorption and scattering coefficients 
given by \cite{laor93} for dust grains of different dimensions, weighted with 
the standard MRN distribution \citep{mathis77}. Grains dimensions range from 0.005 
to 0.25 \mums for graphite, and 0.025 to 0.25 \mums for silicate.

The dust density within the torus is modeled in such a way to allow a gradient along 
both the radial and the angular coordinates: 
\begin{equation}
\label{eqn:density}
\rm \rho(r,\theta)=\alpha \cdot r^\beta \cdot e^{-\gamma |\cos(\theta)|}
\end{equation}
with $\beta$ taking the discrete values 0.0, -0.5 and -1.0, and $\gamma$ the values 0.0 and 6.0.
When $\gamma \ne 0.0$ the dust distribution in no longer a ``flared disk'' but takes
a shape that resembles that of a donut, hence the name ``torus''.
The ``zero value'' $\alpha$ is defined by the value of the equatorial optical
depth at 9.7 micron, ${\rm \tau}_{9.7}$.
One of the novelties of this work is the use of low optical depth tori
(${\rm \tau}_{9.7} < 1$). Even though they are part of model sets present
in the literature (e.g. \citealt{granato94}), they have not been used to
explain the IR emission of AGN. The implications of this approach are explained
in Section \ref{discuss}.

The global model SEDs are computed at different angles of the line-of-sight with 
respect to the torus equatorial plane, in order to account for both type-1 and type-2 
objects emission, and include three contributions: emission from the AGN (partially 
extinguished if the torus intercepts the line-of-sight), thermal and scattering emission 
by dust in each volume element.

Torus models with $\rm R_{out}/R_{in}$ of 300 are a priori excluded from the runs,
even though they belong
to the ``standard'' grid of models presented by \cite{fritz06}, as they imply tori with
physical sizes of several hundred parsecs, sometimes even kpc. For such a ratio, an AGN of 
10$^{46}$ erg/sec accretion luminosity and an inner radius of 1.3 pc would have an outer
radius of $\sim$400 pc. In the general case, only models with $\rm R_{out}/R_{in}$ of 30 
and 100 are allowed, but models with $\rm R_{out}/R_{in}$ of 300 will be revisited
in Section \ref{sec:mips} in order to address specific cases.

\subsection{The stellar component}
\label{sec:stars}

The stellar emission will play a minor role in this work, since we are
dealing with bright, mainly high redshift quasars, for which the galaxy
light is not significant. However, for completeness, we add a stellar
component modeled as the sum of spectra of Simple Stellar Population (SSP) 
models of different age, all assumed with a common (solar) metallicity. The set of SSP used
for this analysis is built with Padova evolutionary tracks \citep{bertelli94}, a Salpeter
IMF with masses in the range
0.15 -- 120 M$_\odot$ and the \cite{jacoby84} library of observed 
stellar spectra in the optical domain. The extension to the UV and IR range is obtained by
means of Kurucz theoretical libraries. Dust emission from circumstellar envelopes 
of AGB stars has been added by \cite{bressan98}. 

\subsection{The starburst component}
\label{sec:sb}

For the cold dust component, the major contributor to the emission at wavelengths
longer than $\sim 30$ \mum, we use two observational starburst templates,
namely M82 as a representative of a ``typical'' starburst IR emission and Arp 220
as representative of a very extinguished starburst.
A more exhaustive approach would require us to provide a physical model of the starburst 
component, which however is far beyond the scope of this work.
Starburst templates are used only when there are observed datapoints
longward of 24 \mums restframe, (typically only when 70 and/or 160 \mums data
are available), and if a torus model fails to
provide an acceptable description of the observational SED.

This choice is, in fact, arbitrary as nothing forbids the presence of a starburst 
component fainter than the 70 \mums detection limit. The flux at 24\mum, though, is
dominated by the torus: the starburst contribution at this wavelength is minimal,
as it coincides with the presence of a deep absorption feature in their SEDs
(see Section \ref{sec:mips}), an with no more datapoints redwards of 24 \mums it is
simply impossible to constrain that part of the SED.
Adding a starburst component in order to fit all objects would only
increase the degeneracy and uncertainties of our results.
Quantities such as the IR luminosity (see Sections \ref{sec:physics}
and \ref{sec:lir}), though, might be seen as lower limits for the objects with no
MIPS 70 and/or 160 \mums detections.

\section{SED FITTING}
\label{sedfit}

Given the large amount of data available, we developed a fully automatic fitting procedure, where
the goodness of the fit is measured in terms of a $\chi^2$ function:
\begin{equation}\label{eqn:chi}
\chi^2=\sum^N_i\left(\frac{O_i-M_i}{\sigma_i}\right)^2
\end{equation}
where $O_i$ are the observed values, $M_i$ the values computed from the model and $\sigma_i$ are the
observed errors of the {\it i-th} photometric point. The expected values from the model
are computed by convolving the synthetic flux with the filters' response curves, after an opportune
normalisation and K-correction is applied.
The dominant component in the UV and NIR (rest-frame) is the accretion disk. 
For very low redshift objects light from the host galaxy might also be present.
The former is clearly distinguishable from a typical stellar SED, 
because it is in general bluer and flat over the entire range of overlap. Hence, if a good fit is not
achieved, stars are removed from the final SED, and a pure AGN component is used at these wavelengths.
Since we are dealing with an AGN sample (i.e. we expect an AGN component to be present in the
observed SEDs of all the objects), the starburst emission, which dominates over the other
components only in the FIR, is included only if there are observed
70 and/or 160 \mums datapoints. 

\begin{figure*}
\centerline{
\psfig{file=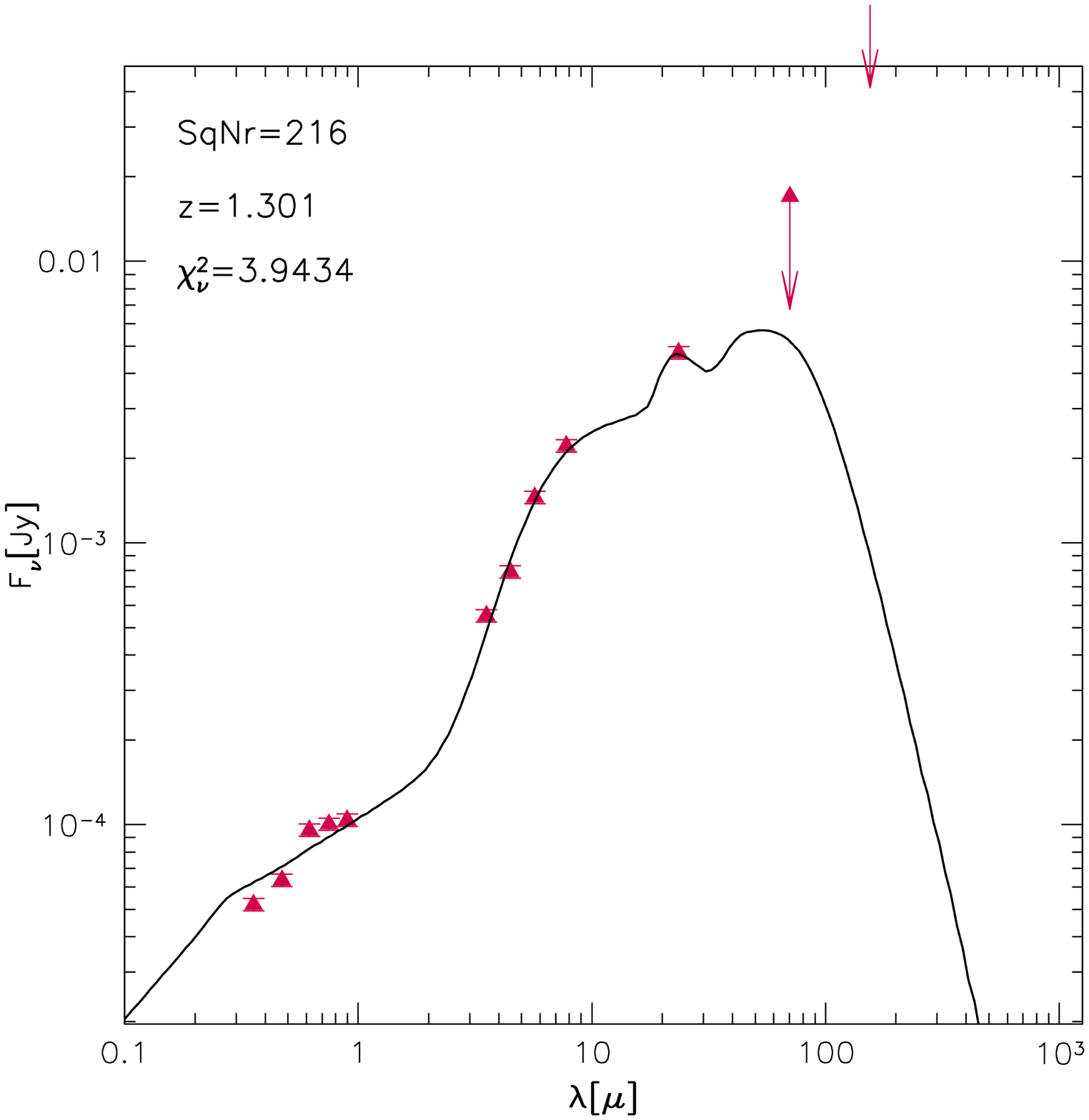,width=6cm}
\psfig{file=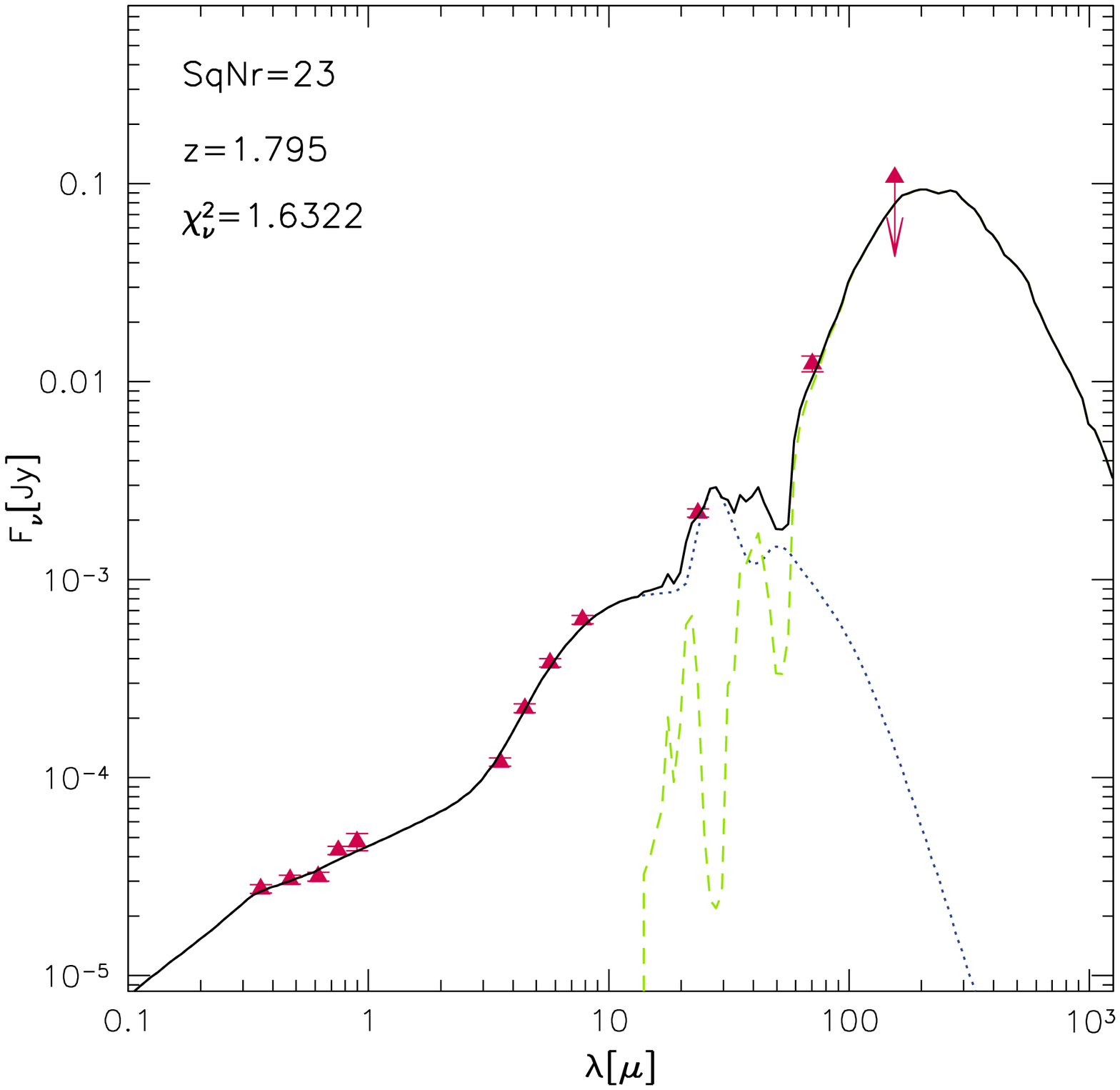,width=6cm}
\psfig{file=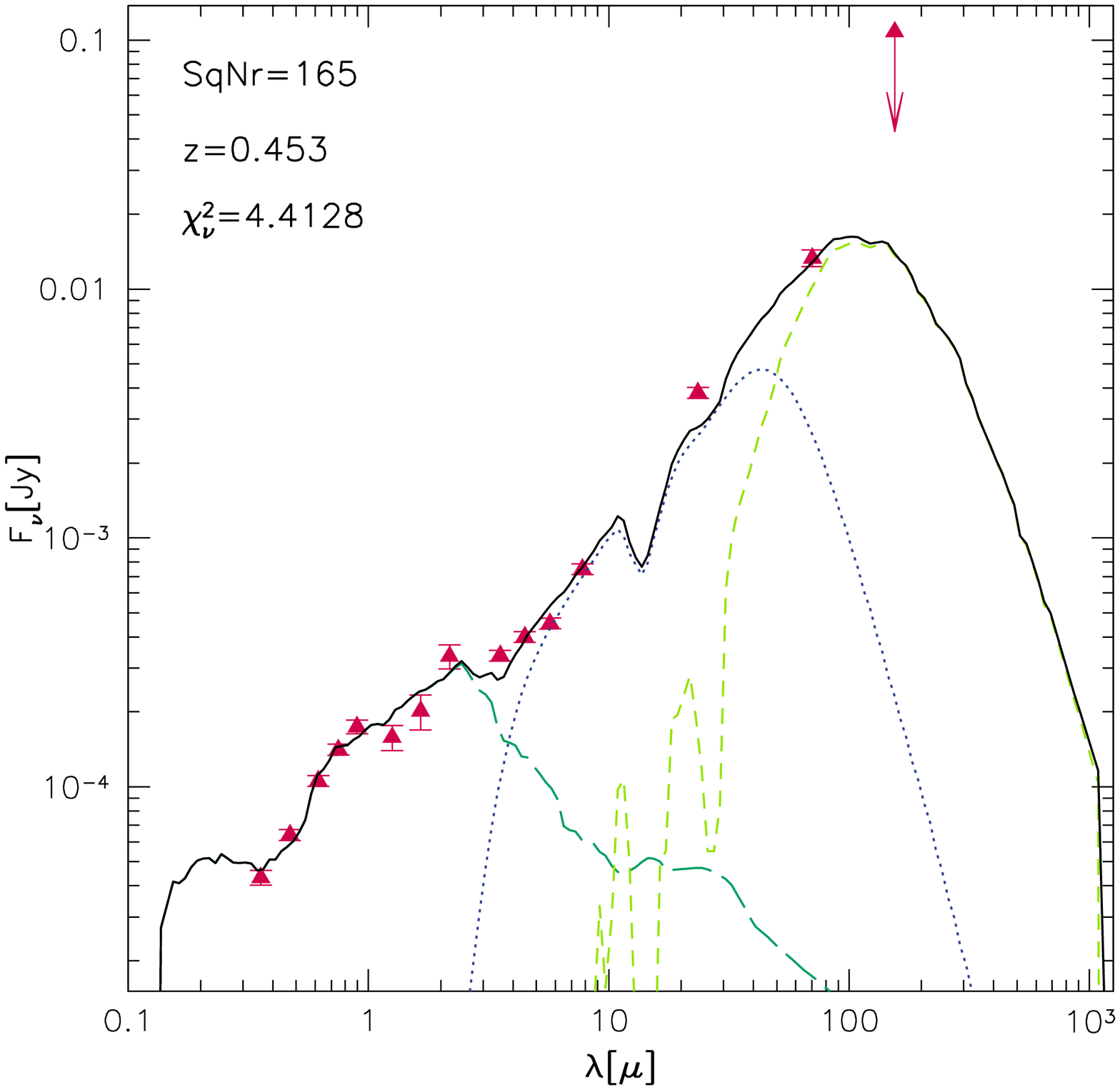,width=6cm}}
\caption{Examples of fits: an object whose SED was
reproduced by an AGN component only (left); a case of an AGN with starburst emission (middle);
and an object with all three components (torus, starburst and stars) present (right).}
\label{fig:fits}
\end{figure*}

Examples of fits are shown in Fig. \ref{fig:fits}: on the left, an object whose SED was
reproduced by an AGN component only; in the middle a case of an AGN with starburst emission;
and in the right an object with all three components (torus, starburst and stars) present. 

Even though the minimum $\chi^2$ will define the best fit, the associated probabilities can
not be taken at face value, as for a number of reasons the derived reduced $\chi^2$s are 
overestimated. First of all, in order to compute the model
magnitudes, the models are convolved with the filters' transmission curves. If the model
is a very accurate representation of the real SED of an object, this convolution will
lead to more accurate results. In our case, however, the optical/UV part of the model
is simply a power law, as already mentioned in Section \ref{sec:tori}, while the SDSS
photometry, that corresponds to the same part of the SED is quite sensitive to the
presence of broad emission lines or the presence of the small blue bump\footnote{The
optical/UV continuum, however, is very well constrained by the available datapoints and
therefore the computation of the accretion luminosity is not affected - see also Section
\ref{sec:physics}}.
Also, as a general remark, the photometric errors are very small (typically of
the order of few percent, as seen in Tables \ref{tab:ugriz} and \ref{tab:jhkspitzer})
for both SDSS and SWIRE
datapoints. And even though we use the errors in the catalogues to properly weight the fits, 
in many cases the computed values of the reduced $\chi^2$s are very high due to the small
photometric errors. In order to avoid excessively high weighting and high $\chi^2$
values, many SED fitting codes impose minimum flux errors (e.g. HyperZ, \citealt{bolzonella00};
ImpZ, \citealt{babbedge04}), here however we chose not to adopt this approach.

\subsection{From SED fitting to physical parameters}
\label{sec:physics}

Based on the best-fit model parameters a number of other physical 
parameters can be derived:
\begin{itemize}
\item the accretion luminosity, $\rm L_{acc}$; this is, the soft X-ray, UV and optical luminosity coming 
from the accretion disc, which provides the main source of dust heating in the torus
\item the IR luminosity, $\rm L_{IR}$, defined as the integral of all the components
in the interval 1 to 1000 \mum
\item the relative contribution of AGN and starburst activity to the IR luminosity 
(the latter is computed only where 70 and/or 160 \mums data are available),
i.e. the fraction of each component with respect to the total IR luminosity
\item the innermost radius of the torus, $\rm R_{in}$, which is the distance at which the 
grains reach their sublimation temperature, averaged over all sizes of graphite grains:
\begin{equation}\label{eqn:rin}
\rm R_{in}\simeq 1.3 \cdot \sqrt{\rm L_{acc}}\cdot T_{1500}^{-2.8} \qquad [pc] ,
\end{equation}
where $\rm T_{1500}$ is the sublimation temperature of the dust grain given in
units of $1500$ K \citep{barvainis87}
\item the optical depth (or extinction in the V band) along the line of sight
\item the hydrogen column density along the line of sight (note that in this case the Galactic
dust-to-gas ratio is implicitly assumed, as a consequence of the use of the MRN distribution
function)
\item the torus full opening angle; this parameter also defines the covering factor, 
which is the percentage of the 
solid angle which is covered by dust in the torus, as seen from the nucleus
\item the mass of dust, $\rm M_{Dust}$, within the torus, which is the integral of all dust 
grains over all volume elements
\end{itemize} 

\subsection{High and low optical depth, ${\rm \tau}_{9.7}$, models}

We also address the issue of low optical depth tori,
namely tori with equatorial ${\rm \tau}_{9.7} \le 1.0$, equivalent to Av $\le 22$.
The obscuring medium (torus) is usually considered to be optically
thick but there are no physical arguments against the existence of optically thin
tori. 
We therefore test the possibility of quasars seen also through low optical depth tori,
as opposed to the ``traditional'' picture of them seen uniquely on lines of sight not 
intercept by the torus. In order to do so, we run the SED fitting twice, once allowing 
for all optical depths, and once allowing only for models with
(${\rm \tau}_{9.7} \ge 1.0$), and compared the results.

\section{RESULTS}
\label{results}

From the 278 quasars comprising the sample, 247 have IRAC coverage, 86 of which have additional 
$JHK$ data from 2MASS (of a total of 97 quasars of our sample detected by 2MASS) and a 
total of 268 has a 24 \mums datapoint. If one requires a good sampling of the SEDs,
one might require detections in at least 8 bands: $u,g,r,i,z$, two out of four 
IRAC bands (due to slight differences in the rastering, objects that lie at the edges of 
the fields might escape detection in 2 of the 4 bands) and MIPS 24 \mum. Similar considerations apply in that
the objects that were not detected in this band were not dropouts but simply happened
to lie outside the covered areas. This condition 
is fulfilled by 237 of the 278 objects. For 70 of these there are 70 \mums detections.
The 41 objects that do not have IRAC data or 24 \mums detections were included in the
sample but  it will be made clear whenever their exclusion causes changes in the results 
mentioned below.

\begin{figure}
\centerline{
\psfig{file=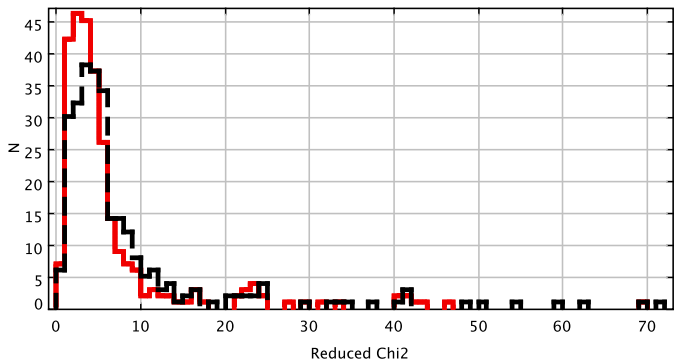,width=10cm}}
\caption{Minimum $\chi^2$ distribution in the case all models are allowed in the
fits (red plain line) and that of only high ${\rm \tau}_{9.7}$ tori models allowed (black dashed line).}
\label{fig:chi2histo}
\end{figure}

Figure \ref{fig:chi2histo}
shows the minimum $\chi^2$ distribution in the case all models are allowed in the
fits (red plain line) and that of only high ${\rm \tau}_{9.7}$ tori models allowed (black dashed line). 
For the statistical analysis of the sample and based on the discussion presented in
Section \ref{sedfit}, we consider the objects for which 
the best fit solutions have reduced $\chi ^2$ lower than 16 for both runs. This
cut is of course arbitrary and it has been chosen because it corresponds to a 
minimum in the $\chi ^2$ distribution corresponding to the run where all ${\rm \tau}_{9.7}$ 
where allowed (see Fig. \ref{fig:chi2histo}). This cut
will exclude in both runs 10\% of the objects. Note that the $\chi ^2$ is
narrower in the case where all models are allowed, but broadens
when constraints on the model grid are imposed. Having chosen a more conservative
cut in the values of reduced $\chi ^2$, e.g. 10, would have excluded another 
$\sim$4\% of the fits and would not further affect the results.

From the 250 of the 278 with reduced $\chi^2<16$ , 190 were better matched 
with an AGN component alone, 46 required an additional starburst component, five more were 
sets of AGN, starburst and stellar component, and another nine
were just composed of stars and an AGN.

Tori with $\beta$ equal to -1.0, were slightly favoured, with 40\% of the observed SEDs
being reproduced by such models, with $\beta$=-0.5 
and 0.0 providing better fits to 32\% and 28\% of the objects, respectively.
Models with $\gamma$ equal to 6.0 (see equation \ref{eqn:density}) provided better fits to 46\% 
of the sample while there was a slight tendency towards
tori with $\rm R_{out}/R_{in}$ of 100 (59\% of the objects). However, when high 
$\rm \tau_{9.7}$ were forced to the fit, the tendencies were reversed, with
56\% of the objects favouring models with $\gamma$ equal to 6.0 and $\rm R_{out}/R_{in}$
of 30 and 50\% of the objects better matching $\beta=-1.0$. All these changes
tend to produce dust more compactly distributed than in the configuration where all
$\rm \tau_{9.7}$ models were permitted and this change occurred most probably in order to
avoid large amounts of IR emission to be produced otherwise by the model.

\begin{figure}
\centerline{
\psfig{file=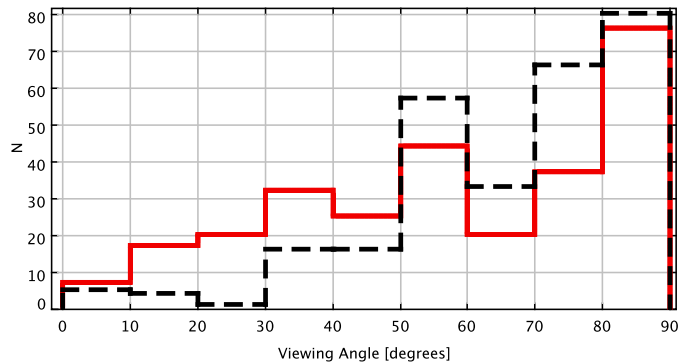,width=10cm}}
\caption{Distribution of viewing angle (in degrees), from the equatorial plane.
The red solid line (black dashed line) corresponds to the run with all tori models allowed
(only high equatorial optical depth, $\rm \tau_{9.7}$, models allowed).}
\label{fig:angle}
\end{figure}

Fig. \ref{fig:angle} shows the distribution of the best values for the viewing angles,
measured from the equatorial plane towards the pole, for objects with minimum $\chi^2$s 
below the selected threshold. The colour coding is the same as in Fig. \ref{fig:chi2histo}.

When all optical depths are allowed, the SEDs of more objects could be reproduced with
configurations allowing for the line of sight (LoS) to intercept a (low $\rm \tau_{9.7}$)
torus. When only high $\rm \tau_{9.7}$ values are allowed, the histogram of viewing angles
is depleted towards the low angles end and boosted towards the large angles end, as
expected from our current view of the AGN paradigm, in which type-1 objects can not be
seen through the torus. In both cases, however, for the majority of objects the viewing
angle is larger than half the torus effective angle, and therefore the LoS does not
intercept it.

\subsection{The covering factor}

\begin{figure}
\centerline{
\psfig{file=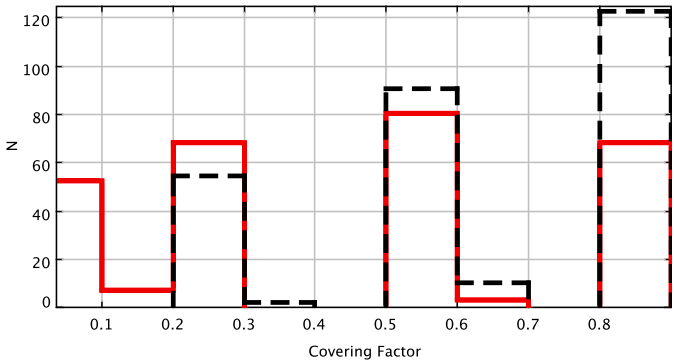,width=10cm}}
\caption{Distribution of the best fit values for the torus CF.
Colour-coding is the same as for Figs. \ref{fig:chi2histo} and \ref{fig:angle}.
The gaps are due to the discrete values given to the torus half opening angle 
when the grid of models was built (see text fro more details).}
\label{fig:histoCF}
\end{figure}

The covering factor (CF), as already mentioned, is the percentage of the solid
angle blocking the light of the AGN due to dust absorption. For models
with high equatorial optical depths, $\rm \tau_{9.7}$, it is proportional
to the cosine of the effective opening angle of the torus.
Fig. \ref{fig:histoCF} shows the distribution of CF
for the run with all models allowed and that for which only high values 
of $\rm \tau_{9.7}$ were used. The gaps are due to the discrete values given 
to the torus half opening angle when the grid of models was built, namely of 20, 40 and 60
degrees. 
For low $\rm \tau_{9.7}$ models and models with density decreasing towards high
altitudes above the equatorial plane ($\gamma > 0$ in equation \ref{eqn:density})
though, the covering factor has to be recomputed to account for the
fact that the torus obscuration is considered effective only where
$\tau(\lambda=0.3{\rm \mu})$ $>1$.

The colour coding is the same as that in Fig. 
\ref{fig:chi2histo}.
The plot implies that very low optical depth tori correspond to very small
covering factors (first bin of red solid histogram). This happens because in 
these cases the CF is recomputed to account for the 
fact that the quasars could be seen through an optically very thin medium. We
consider that the LoS intercepts the torus when the optical depth at 3000\AA \, reaches
the value of 1. When these models are excluded, the CF assumes
a steeper distribution, with many more objects having much larger dust coverage.

\subsection{IR luminosity and mass of dust}
\label{sec:lir}

For objects whose MIR SED can be represented by a torus component only, the
ratio of $\rm L_{acc}$ over the IR luminosity, $\rm L_{IR}$, would be another
indicator of the obscuration. Infrared radiation is UV/optical radiation
reprocessed by the dust composing the torus. For type-1 objects, for which the
central engine is directly observed, this ratio is expected to take values
larger than unity, as indeed is the case.
Furthermore, $\rm L_{acc}/L_{IR}$ does not show any dependence neither on 
redshift nor on $\rm L_{acc}$.
\begin{figure}
\centerline{
\psfig{file=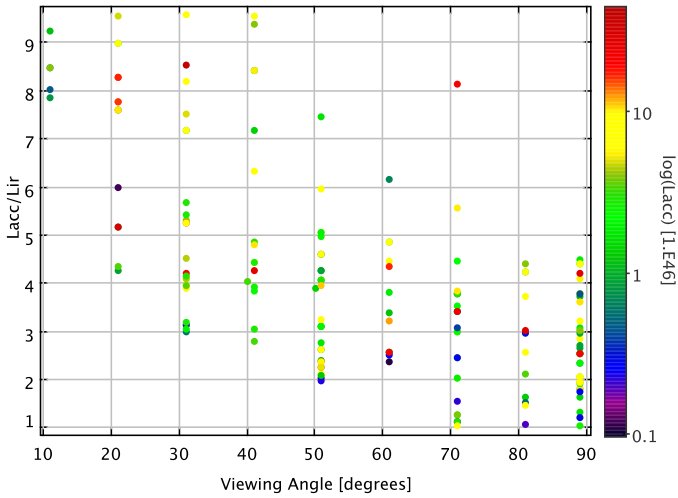,width=10cm}}
\caption{$\rm L_{acc}/L_{IR}$ as a function of the viewing angle and the
accretion luminosity, $\rm L_{acc}$.}
\label{fig:LaccLir}
\end{figure}
There is, however, a clear tendency of decreasing $\rm L_{acc}/L_{IR}$ with
increasing viewing angle shown in Fig. \ref{fig:LaccLir}, 
with a factor of 4 of difference between 10$^{\rm o}$ and 90$^{\rm o}$.
This tendency is due to the fact that $\rm L_{IR}$ is orientation-independent
while the measurements of $\rm L_{acc}$ are not, translated into a constant 
$\rm L_{IR}$ and a $decreasing$ $\rm L_{acc}$ with increasing viewing angle.
From the results of the best fit models one notes that objects with
best viewing angles of less than 30$^{\rm o}$, are also assigned models
with an angular variation of the dust density, i.e. $\gamma=-6.0$. The ratio
of $\gamma(-6.0)/\gamma(0.0)$ drops to 0.85 for objects with best viewing 
angles between 40$^{\rm o}$ and 60$^{\rm o}$ and to 0.5 for objects
seen from even higher viewing angles. The objects with $\gamma=-6.0$ tend to have,
on average, larger accretion luminosities suggesting that $\rm L_{acc}$ might be
influencing the dust distribution as well as the torus opening angle.

\begin{figure}
\centerline{
\psfig{file=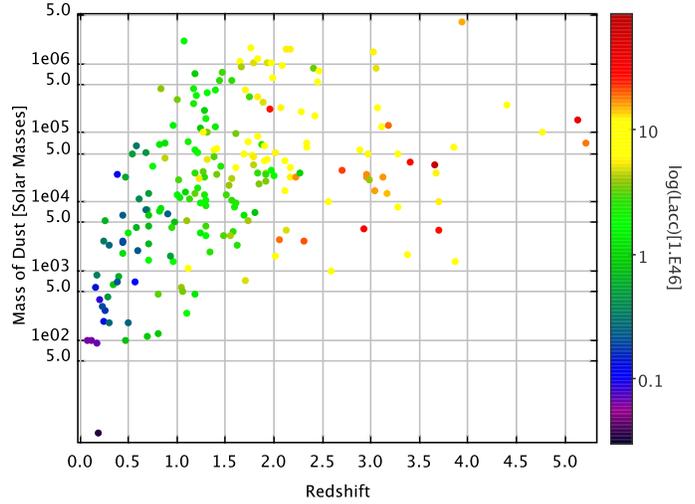,width=10cm}}
\caption{Mass of dust, $\rm M_{Dust}$, as a function of redshift, $z$, and accretion
luminosity, $\rm L_{acc}$.}
\label{fig:MdustZ}
\end{figure}

$\rm L_{IR}$ also increases with
increasing $\rm M_{Dust}$, as larger amounts of dust produce stronger IR emission.
The mass of dust, $\rm M_{Dust}$, increases with redshift
as well as with $\rm L_{acc}$ (Fig. \ref{fig:MdustZ}).
The dependence on the redshift, however, is mainly an observational bias: $\rm M_{Dust}$ is computed
integrating over all dust grains over all volume elements of the torus and therefore depends
on R$_{\rm in}$ that scales with $\sqrt (\rm L_{acc})$ (from equation 
\ref{eqn:rin}), which increases with increasing redshift (Scott effect). 

In order to make sure all tendencies with redshifts are simply the result of
the observational bias all flux limited samples suffer from, we divided part of the sample
to slices of $\rm L_{acc}$ and redshift and computed the mean for $\rm M_{Dust}$,
$\rm L_{IR}$ and CF in these bins. The bins were chosen in a way to have at least 3 redshift
ones per luminosity range. The result are given in Table \ref{tab:bins}.
The three quantities show no statistically significant trend with redshift in either
of the two $\rm L_{acc}$ bins. The average values of the three quantities show an increase
in lower luminosity bin but can be considered constant within the errors. 

\begin{table*}
\caption{$\rm M_{Dust}$, $\rm L_{IR}$ and CF averaged over redshift and luminosity bins.}
\small
\begin{tabular}{c c | c c | c c c}
\hline
\hline
bin \# & N & $\rm L_{acc}$           & $z$ & $\rm M_{Dust}$  & $\rm L_{IR}$            & CF \\
       &   & $\rm [10^{46} erg/sec]$ &     & $\rm M_{\odot}$ & $\rm [10^{46} erg/sec]$ &    \\ 
\hline
\hline
1 &  8 & 2.0--5.0  & 0.5--1.0 & 0.66$\pm$1.34 & 0.96$\pm$0.35 & 0.42$\pm$0.34 \\
2 & 37 & 2.0--5.0  & 1.0--1.5 & 1.22$\pm$1.75 & 1.13$\pm$0.91 & 0.43$\pm$0.27 \\
3 & 27 & 2.0--5.0  & 1.5--2.0 & 1.75$\pm$2.99 & 1.25$\pm$0.71 & 0.49$\pm$0.27 \\
\hline
4 &  7 & 5.0--10.0 & 1.0--1.5 & 2.59$\pm$3.95 & 1.97$\pm$1.61 & 0.31$\pm$0.19 \\
5 & 18 & 5.0--10.0 & 1.5--2.0 & 3.88$\pm$4.83 & 2.90$\pm$2.02 & 0.52$\pm$0.26 \\
6 & 14 & 5.0--10.0 & 2.0--2.5 & 2.84$\pm$4.50 & 2.44$\pm$1.24 & 0.45$\pm$0.28 \\
\hline
\hline
\end{tabular}
\label{tab:bins}
\end{table*}

The results on the accretion luminosity and the dust mass are
independent of the choice to impose high optical depth models in the fits.
The distributions shown in Fig. \ref{fig:LaccLir} are indistinguishable.
This confirms the robustness of the derived
values of $\rm L_{acc}$, as further discussed in Section \ref{degeneracies}.

\subsection{The torus dimensions}

The inner radius of the torus, i.e. the minimum distance at which dust grains
can exist, is computed from equation \ref{eqn:rin} and for the sample under study
typically takes values up to 5 pc, 
with a broad distribution and a peak at $\sim$2 pc. 
Equation \ref{eqn:rin}, however, only takes into account the accretion luminosity 
as a heating source for the dust. And while this is realistic for the majority
of cases, there are some particular conditions, for which the thermal emission from dust 
itself can become an important, non-negligible contributor to its self-heating.
This this likely to occur in models with
steep density profiles ($\beta \le -1$) and high optical depths
that will give rise to so high density at the inner radius
that a large amount of the dust-emitted radiation is re-absorbed locally and the
sublimation temperature is reached and exceeded for all the dust grains.
When this happens during the computation of a model, 
the inner radius is moved to larger values, at steps of 0.2 pc, 
and the calculation of the equilibrium temperature starts again. This procedure is 
repeated until the temperature at the innermost radius remains below the sublimation value.

As already mentioned, our models use a fixed R$_{\rm out}$/R$_{\rm in}$ ratio, namely of 30
and 100. These cuts are only useful when the dust density is assumed constant and does 
not vary neither with the distance to the centre nor with the angle from the equatorial
plane, i.e. when $\beta=0.0$ and $\gamma=0.0$. When $\beta=-1.0$ though, the dust density
rapidly decreases with the distance to the centre and the value of R$_{\rm out}$ is in practice
much smaller than that given by the constant R$_{\rm out}$/R$_{\rm in}$ ratios. We find the outer radii of
the tori to be typically some tens of parsecs 
large but can reach, in a few cases, a few hundreds of parsecs.

\subsection{Equatorial optical depth and hydrogen column density, N$_{\rm H}$}

The hydrogen column density, N$_{\rm H}$, is proportional to the
optical depth and is defined as \citep{fritz06}:
\begin{equation}
\rm
 \tau_{eq}(\lambda)=N_H\times\sum_{id=1}^{Ndust}\left[Q_{id}^{A}
(\lambda)+Q_{id}^{S}(\lambda)\right]a^2_{id} \pi \cdot Nd_{id}
\label{eqn:tauNh}
\end{equation}
where Q$_{\rm id}^{\rm A}$ and Q$_{\rm id}^{\rm S}$ are the absorption and scattering coefficients,
of the {\it id}-th dust grain, respectively, N$_{\rm tot}$ is the total number of grain
sizes, a$_{\rm id}$ their radius and N$_{\rm id}$ is the number of grain, of a given
dimension, calculated with respect to hydrogen. The equatorial optical depth,
$\rm \tau_{eq}$, is in general different of that computed along the LoS.

\begin{figure}
\centerline{
\psfig{file=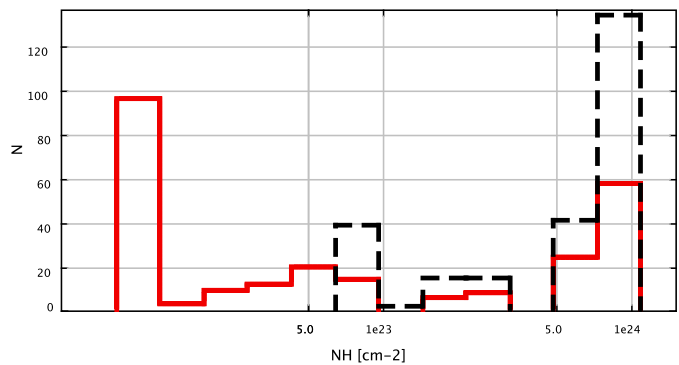,width=10cm}}
\caption{Equatorial hydrogen column density, N$_{\rm H}$, distribution for
all models (red plain line) and high optical depth models (black dashed line).}
\label{fig:Nh}
\end{figure}

Fig. \ref{fig:Nh} shows that high $\rm \tau_{9.7}$ models correspond to very
high equatorial hydrogen column densities, N$_{\rm H}$, while allowing for
low $\rm \tau_{9.7}$ values of N$_{\rm H}$ as low as a few times 10$^{21}$ are
feasible. In the latter case, the N$_{\rm H}$ distribution is bimodal
peaking at the extreme of the values allowed.

The N$_{\rm H}$ values as derived from the model may not be directly comparable 
to observed ones (e.g. those computed via X-ray observations). In the present study
we assume a Galactic dust-to-gas ratio, that is likely to vary from one galaxy to 
another. Furthermore, the model value is only accounting for hydrogen within the 
torus, while observations one takes into account also the gas within the torus 
inner radius, as well as the gas likely present in the outer parts of the galaxy 
as, e.g., diffuse medium.

\subsection{The objects with 70 and 160 \mums detections}
\label{sec:mips}

\begin{figure}
\centerline{
\psfig{file=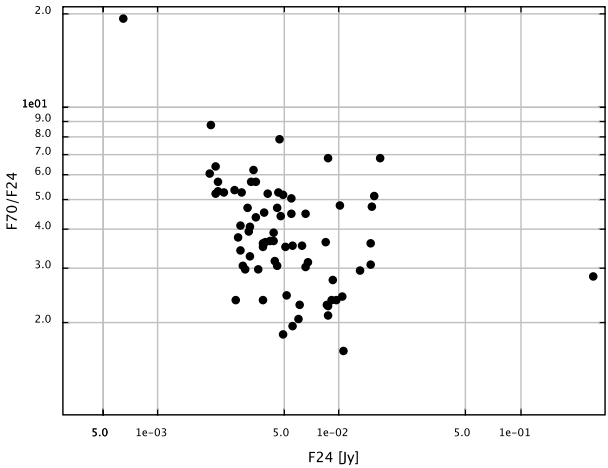,width=9cm}}
\caption{Observed F$_{70}/$F$_{24}$ versus 24 \mums flux for the 70 objects with 
70 \mums detections.}
\label{fig:f70f24}
\end{figure}

The objects with 70 \mums detection represent 25\% of the sample under study.
They have already been included in our previous analysis but are worth
a closer scrutiny. Their 70 and 160 \mums fluxes, as well as their properties, derived from
the best fit models are shown in Table \ref{tab:70usample}. 
The distribution of their F$_{70}/$F$_{24}$ colour versus their 24 \mums fluxes,
F$_{24}$, is shown in Fig. \ref{fig:f70f24}. F$_{70}/$F$_{24}$ usually
takes values in the range between 1 and 8, even for the brightest 24 \mums sources,
depending on the relative contribution of the starburst to the IR luminosity.
For objects with F$_{24} \le 0.3$ mJy there are no 70 \mums detections because of the
70 \mums limiting flux of the survey and the 70 \mums source definition (see Section \ref{data}).

\begin{figure*}
\centerline{
\psfig{file=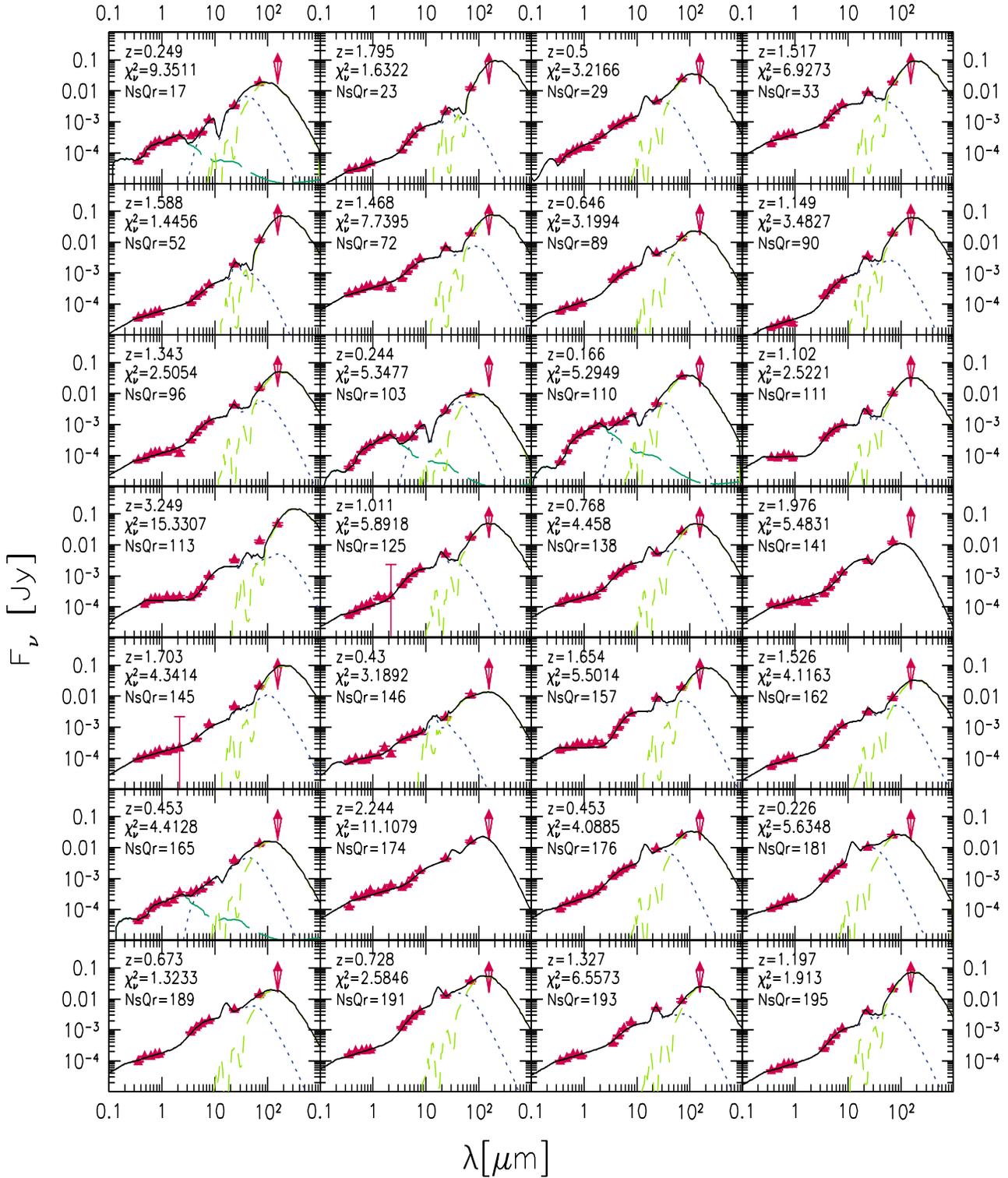,height=.90\textheight,width=.97\textwidth}}
\caption{Observed (red triangles) and model (torus: blue dotted line; starburst:
dashed light green; stars: long-dashed dark green)
SEDs for the 55 objects with 70 \mums detection and minimum $\chi^2 < 16$.}
\label{fig:qsosbseds1}
\end{figure*}

\begin{figure*}
\centerline{
\psfig{file=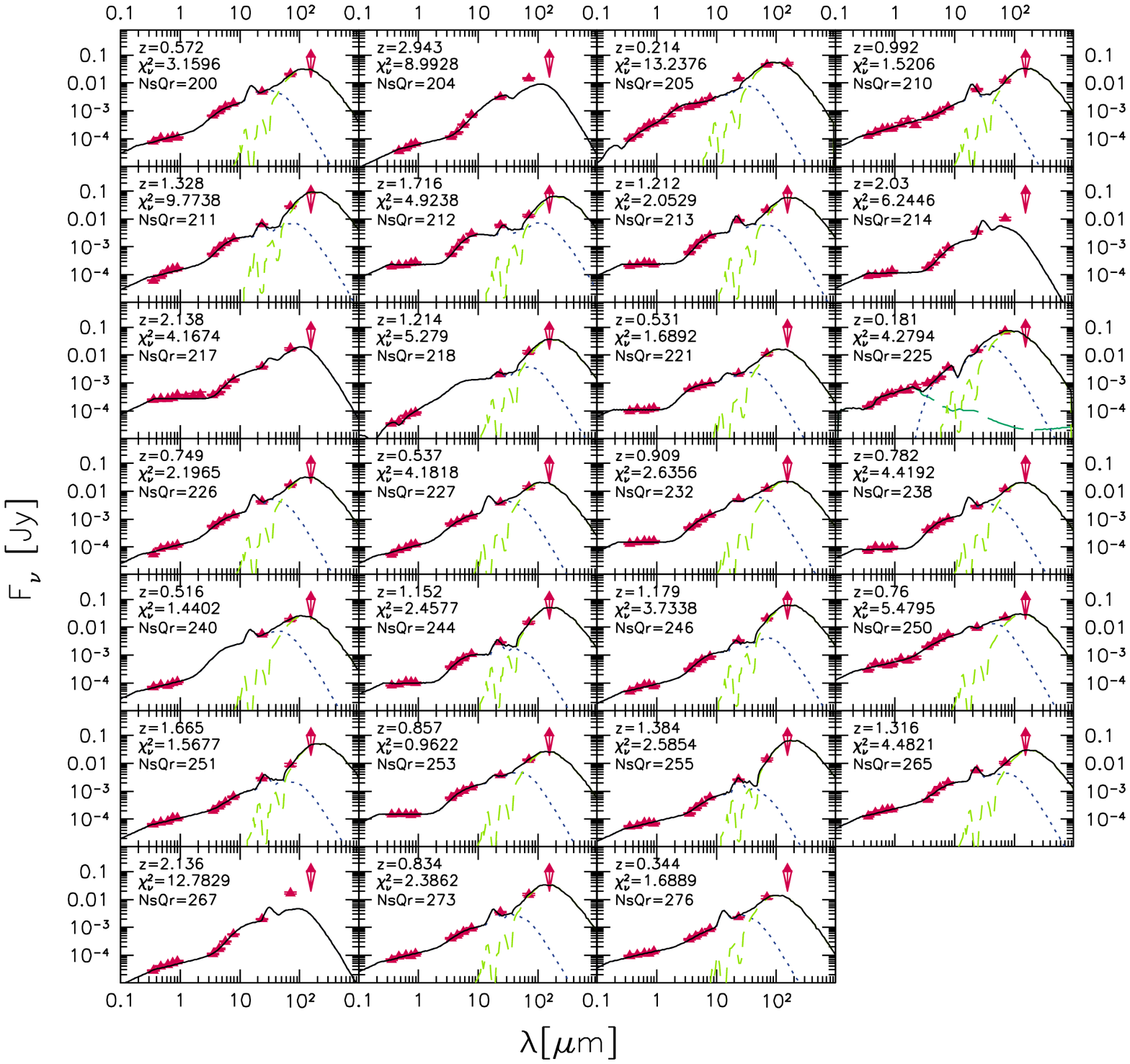,height=.90\textheight,width=.97\textwidth}}
\caption{Same as in Fig. \ref{fig:qsosbseds1}.}
\label{fig:qsosbseds2}
\end{figure*}

Of the 70 objects, 55 had acceptable fits (with 
$\chi^2 \le 16$), with only one detected at 160 \mum.
Five of the 55 observed SEDs were 
reproduced with a single component only, namely a torus; the other 50
required a starburst template. Five low redshift objects
also required a stellar population to fit the optical/UV part 
of their SEDs. As mentioned in Section \ref{sec:sb} two starburst
templates were used in the fits, Arp220 and M82. 
Figs. \ref{fig:qsosbseds1} and \ref{fig:qsosbseds2}
shows the observed SEDs (red triangles) and
the components of the models reproducing them for the 55 objects with good 
reduced $\chi^2$ and an assigned starburst. The torus components are shown in
blue dotted lines, the starbursts in light green dashed lines and the stellar
components in dark green long-dashed lines. The total models are illustrated
in solid black lines.

\begin{table*}
\small
\begin{tabular}{r c c c c c r c}
\hline
\hline
SqNr & RA & Dec & z & S$_{70}$ [mJy] & S$_{160}$ [mJy] & SB & AGN$_{frac}$\\
\hline
\hline
  17  & 10:34:21.24 & +58:06:53.0 & 0.249 & 18.71$\pm$1.33 & --		     & Arp220  & 0.3939   \\
  23  & 10:35:52.62 & +58:43:25.8 & 1.795 & 12.35$\pm$1.13 & --		     & Arp220  & 0.2059   \\
  29  & 10:36:51.94 & +57:59:51.0 & 0.5   & 24.12$\pm$1.2  & --		     & Arp220  & 0.5809   \\
  33  & 10:37:24.74 & +58:05:13.1 & 1.517 & 19.68$\pm$0.95 & --		     & Arp220  & 0.5406   \\
  37  & 10:38:03.35 & +57:27:01.7 & 1.285 & 80.51$\pm$1.53 & 130.25$\pm$5.0  & Arp220  & 0.319    \\
  52  & 10:40:49.10 & +59:06:24.4 & 1.588 & 11.69$\pm$0.83 & --		     & Arp220  & 0.1908   \\
  53  & 10:40:58.80 & +58:17:03.3 & 0.071 & 17.04$\pm$0.93 & --		     & Arp220  & 0.723    \\
  60  & 10:41:19.23 & +57:45:00.1 & 0.067 & 705.5$\pm$6.94 & 642.91$\pm$6.61 & M82     & 0.0015   \\
  65  & 10:41:55.16 & +57:16:03.3 & 1.721 & 25.44$\pm$1.1  & --		     & Arp220  & 0.2639   \\
  72  & 10:42:55.65 & +57:55:49.9 & 1.468 & 19.55$\pm$0.92 & --		     & Arp220  & 0.5584   \\
  89  & 10:45:26.72 & +59:54:22.5 & 0.646 & 13.92$\pm$1.19 & --		     & Arp220  & 0.6861   \\
  90  & 10:45:39.27 & +58:07:11.1 & 1.149 & 19.55$\pm$1.27 & --		     & Arp220  & 0.2901   \\
  96  & 10:46:22.61 & +57:40:34.2 & 1.343 & 15.11$\pm$1.15 & --		     & Arp220  & 0.5129   \\
  103 & 10:47:05.57 & +58:27:41.8 & 0.244 & 9.699$\pm$0.73 & --		     & Arp220  & 0.3954   \\
  108 & 10:47:39.50 & +56:35:07.2 & 0.303 & 59.04$\pm$1.13 & 376.88$\pm$0.01 & Arp220  & 0.2771   \\
  110 & 10:47:56.98 & +56:24:47.2 & 0.166 & 36.72$\pm$1.38 & --		     & Arp220  & 0.2677   \\
  111 & 10:47:57.52 & +57:34:51.8 & 1.102 & 10.39$\pm$0.89 & --		     & Arp220  & 0.478    \\
  112 & 10:48:06.19 & +56:30:21.5 & 1.905 & 10.55$\pm$1.44 & --		     & Arp220  & 0.3462   \\
  113 & 10:48:09.18 & +57:02:42.0 & 3.249 & 13.19$\pm$0.92 & 47.83$\pm$4.5   & Arp220  & 0.3071   \\
  123 & 10:49:25.88 & +57:54:21.6 & 0.071 & 21.00$\pm$1.05 & --		     & Arp220  & 0.8394   \\
  125 & 10:49:30.46 & +59:20:32.6 & 1.011 & 17.76$\pm$0.95 & --		     & Arp220  & 0.5055   \\
  138 & 10:51:06.12 & +59:16:25.2 & 0.768 & 27.47$\pm$0.93 & --		     & Arp220  & 0.6253   \\
  141 & 10:51:53.78 & +56:50:05.7 & 1.976 & 12.35$\pm$0.89 & --		     &   --    & 1.0      \\
  142 & 10:51:58.53 & +59:06:52.1 & 1.814 & 24.47$\pm$0.86 & --		     & Arp220  & 0.2643   \\
  145 & 10:52:59.88 & +59:22:34.0 & 1.703 & 21.36$\pm$1.04 & --		     & Arp220  & 0.3807   \\
  146 & 10:53:08.25 & +59:05:22.2 & 0.43  & 10.85$\pm$1.36 & --		     &   --    & 0.4909   \\
  152 & 10:54:04.11 & +57:40:19.8 & 1.101 & 18.23$\pm$0.95 & 436.6$\pm$4.53  & Arp220  & 0.5371   \\
  157 & 10:54:47.29 & +58:19:09.5 & 1.654 & 19.44$\pm$0.95 & --		     & Arp220  & 0.5587   \\
  162 & 10:55:49.88 & +58:26:01.5 & 1.526 & 8.989$\pm$0.75 & --		     & Arp220  & 0.5942   \\
  165 & 10:56:39.42 & +57:57:21.5 & 0.453 & 13.31$\pm$1.02 & --		     & Arp220  & 0.4404   \\
  168 & 10:57:05.41 & +58:04:37.5 & 0.14  & 113.4$\pm$1.25 & 211.53$\pm$5.52 & Arp220  & 0.4732   \\
  171 & 10:57:17.33 & +58:01:03.0 & 3.311 & 12.35$\pm$0.82 & --		     &   --    & 1.0      \\
  174 & 10:59:02.04 & +58:08:48.7 & 2.244 & 16.07$\pm$0.95 & --		     &   --    & 1.0      \\
  176 & 10:59:59.93 & +57:48:48.2 & 0.453 & 25.07$\pm$1.04 & --		     & Arp220  & 0.7305   \\
  181 & 11:02:23.57 & +57:44:36.3 & 0.226 & 24.95$\pm$1.87 & --		     & Arp220  & 0.7661   \\
  184 & 15:56:49.75 & +54:35:51.3 & 1.71  & 8.880$\pm$0.81 & --		     & Arp220  & 0.5604   \\ 
  189 & 16:00:15.69 & +55:23:00.0 & 0.673 & 13.80$\pm$1.04 & --		     & Arp220  & 0.7853   \\
  191 & 16:01:28.54 & +54:45:21.4 & 0.728 & 37.92$\pm$1.25 & --		     & Arp220  & 0.7439   \\
  193 & 16:02:30.04 & +54:36:58.5 & 1.327 & 9.020$\pm$1.04 & --		     & Arp220  & 0.674    \\
  195 & 16:02:50.97 & +54:50:57.9 & 1.197 & 20.87$\pm$1.09 & --		     & Arp220  & 0.3485   \\
  200 & 16:05:23.11 & +54:56:13.4 & 0.572 & 21.00$\pm$1.11 & --		     & Arp220  & 0.6389   \\
  204 & 16:06:37.88 & +53:50:08.4 & 2.943 & 14.64$\pm$0.99 & --		     &   --    & 1.0      \\
  205 & 16:06:55.36 & +53:40:16.8 & 0.214 & 45.72$\pm$1.99 & 50.85$\pm$4.63  & Arp220  & 0.5948   \\
  206 & 16:07:05.17 & +53:35:58.6 & 3.653 & 22.2$\pm$0.8   & --		     &   --    & 1.0      \\	
  210 & 16:09:13.19 & +53:54:29.6 & 0.992 & 12.23$\pm$0.85 & --		     & Arp220  & 0.6318   \\	
  211 & 16:09:43.67 & +53:30:41.0 & 1.328 & 28.92$\pm$1.01 & --		     & Arp220  & 0.441    \\	
  212 & 16:09:50.72 & +53:29:09.5 & 1.716 & 13.68$\pm$1.14 & --		     & Arp220  & 0.5853   \\	
  213 & 16:09:53.03 & +53:34:43.9 & 1.212 & 21.36$\pm$0.94 & --		     & Arp220  & 0.6154   \\	
  214 & 16:10:07.12 & +53:58:14.1 & 2.03  & 10.53$\pm$1.21 & --		     &   --    & 1.0      \\	
  217 & 16:12:38.28 & +53:22:55.1 & 2.138 & 17.28$\pm$1.36 & --		     &   --    & 1.0      \\
  218 & 16:12:57.60 & +52:48:32.9 & 1.214 & 13.31$\pm$1.38 & --		     & Arp220  & 0.5722   \\
  221 & 16:30:31.47 & +41:01:45.8 & 0.531 & 11.35$\pm$1.03 & --		     & Arp220  & 0.6577   \\
  225 & 16:31:28.60 & +40:45:36.0 & 0.181 & 71.40$\pm$1.62 & --		     & Arp220  & 0.4198   \\
  226 & 16:31:35.47 & +40:57:56.4 & 0.749 & 16.92$\pm$0.77 & --		     & Arp220  & 0.6203   \\
  227 & 16:31:43.76 & +40:47:35.7 & 0.537 & 14.15$\pm$1.37 & --		     & Arp220  & 0.6927   \\
  232 & 16:32:25.56 & +41:18:52.6 & 0.909 & 12.47$\pm$1.07 & --		     & Arp220  & 0.7071   \\
  235 & 16:33:08.29 & +40:33:21.4 & 0.404 & 30.47$\pm$0.97 & 103.18$\pm$4.183& Arp220  & 0.396    \\
  238 & 16:33:52.34 & +40:21:15.6 & 0.782 & 11.72$\pm$1.1  & --		     & Arp220  & 0.6571   \\
  240 & 16:34:26.28 & +41:52:15.5 & 0.516 & 19.44$\pm$1.6  & --		     & Arp220  & 0.7376   \\
  244 & 16:35:31.06 & +41:00:27.4 & 1.152 & 15.23$\pm$1.09 & --		     & Arp220  & 0.3848   \\
  246 & 16:36:18.51 & +41:50:58.9 & 1.179 & 20.87$\pm$1.0  & --		     & Arp220  & 0.3819   \\
  247 & 16:36:31.29 & +42:02:42.5 & 0.061 & 53.63$\pm$1.37 & 91.19$\pm$6.05  & Arp220  & 0.7652   \\
  250 & 16:37:09.32 & +41:40:30.8 & 0.76  & 22.55$\pm$0.93 & --		     & Arp220  & 0.8411   \\
  251 & 16:37:14.40 & +41:57:02.1 & 1.665 & 8.930$\pm$1.03 & --		     & Arp220  & 0.3886   \\
  253 & 16:37:26.89 & +40:44:32.9 & 0.857 & 13.56$\pm$0.94 & --		     & Arp220  & 0.6531   \\
  255 & 16:37:59.16 & +42:03:36.8 & 1.384 & 14.15$\pm$1.02 & --		     & Arp220  & 0.2503   \\
  265 & 16:40:18.34 & +40:58:13.1 & 1.316 & 10.76$\pm$0.85 & --		     & Arp220  & 0.7006   \\
  267 & 16:40:39.31 & +40:31:35.1 & 2.136 & 17.04$\pm$1.32 & --		     &   --    & 1.0      \\
  273 & 16:43:11.84 & +40:50:43.3 & 0.834 & 15.11$\pm$1.01 & --		     & Arp220  & 0.5112   \\
  276 & 16:43:43.25 & +40:56:54.4 & 0.344 & 12.11$\pm$1.37 & --		     & Arp220  & 0.6623   \\
\hline
\hline
\end{tabular}
\caption{Coordinates, redshift, 70 and 160 \mums fluxes, starburst template used in the best fit 
and AGN fraction for the 70 objects with 70 \mums detections, ordered by RA.}
\label{tab:70usample}
\end{table*}

All but one objects
were assigned an Arp220-like template to reproduce their 70 \mums detection.
The reason for that is the very bright limiting magnitude of the survey in the
MIPS Ge bands. 
An Arp220-like component could make the 70 \mums bright enough 
to be observed without contributing to the IRAC fluxes while and M82 component
would also contribute considerably to the IRAC and MIPS 24 \mums fluxes.

\begin{figure}
\centerline{
\psfig{file=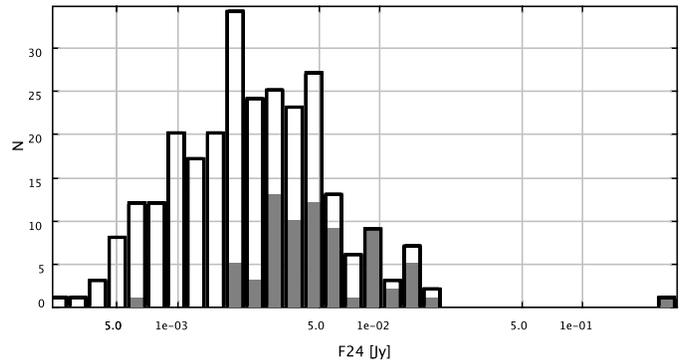,width=10cm}}
\caption{24 \mums flux histogram of the entire sample (open) and of the sources with
70 \mums counterparts (grayed).}
\label{fig:histo24}
\end{figure}

Another point favouring the conclusion that objects not detected at 70 \mums could
in fact be characterised by a less extreme starburst (e.g. M82-like) starburst is the fact that 
even though objects with 70 \mums detections are 
the brightest at 24 \mum, not all bright 24 \mums sources have a 70 \mums detection, 
as seen in Fig. \ref{fig:histo24}. The opposite, i.e. all bright 24 \mums sources 
observed at 70 \mum, would imply that fainter 24 \mums sources
would have different F$_{70}$/F$_{24}$ ratios and therefore a different starburst 
component (M82), but there is no observational evidence in favour. 

\begin{figure}
\centerline{
\psfig{file=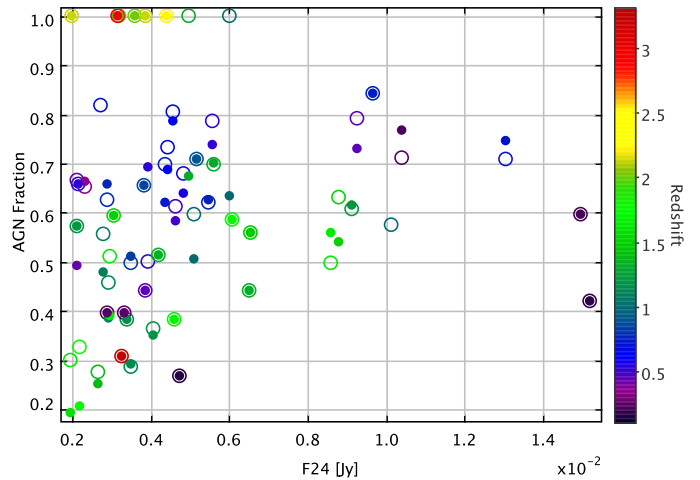,width=10cm}}
\caption{The fraction of IR luminosity, as defined in Section
\ref{sec:physics}, attributed to the AGN as a function of their 24\mums flux
and redshift, for the 55 objects with 70 \mums detections and $\chi^2<16$.
Filled (open) circles correspond to the run with all (high) $\tau_{9.7}$.}
\label{fig:fracagn}
\end{figure}

The contributions of the AGN to the total IR luminosity, as defined in Section
\ref{sec:physics}, for the objects with starburst components
is shown in Fig. \ref{fig:fracagn} (filled circles for the run with all
$\tau_{9.7}$, open circles for the run with high values of $\tau_{9.7}$), 
as a function of their 24 \mums flux and redshift. 
The mean difference of the AGN contribution derived from the two runs is of 3\%
with a standard deviation of 8\%, the results can therefore be considered independent
of the choice of $\tau_{9.7}$s.
The AGN fraction varies from $\sim$20\% up to 100\%, with lower redshift objects
tending to have higher AGN contribution. The increase of the AGN fraction with 
the 24 \mums flux is due to the fact that, when the starburst component is an Arp220,
the 24 \mums flux is produced by the torus alone, as it corresponds to a very deep
absorption feature in the starburst template. This requires a brighter AGN
and a lower starburst component in order to keep the ratio F$_{70}$/F$_{24}$ in the
observed range.
The presence of strong starburst component in many of the cases is an interesting
result, when one considers the fact that the quasars under study are among the
brightest in the sky.

It could be argued that allowing for larger dust distributions
might increase the fraction of the AGN contribution to the
infrared luminosity or reproduce the 70 micron point without the need of a starburst component.
In order to account for this possibility but also in an effort to reproduce the few 
observed 160 \mums datapoints and potentially
improve the fit for the rest of the objects with 70 \mums data, we re-run the SED fitting
allowing for models with $\rm R_{out}/R_{in}=300$, {\it a priori} excluded from the runs
for reasons presented in Section \ref{sec:tori}.
In none of the cases, however, was a torus component alone enough to reproduce the 70 \mums point. 
In fact, all objects required a starburst component (again Arp220) with a contribution within at most 10\%
lower than that predicted when no large tori were allowed. Furthermore,
the 160 \mums points were still not properly reproduced, as all combinations of torus and
starburst templates produced model SEDs lying significantly below the observed
160 \mums points.

\subsection{Multiple local minimae and degeneracy}
\label{degeneracies}

SED fitting in a multi-parameter space usually involves
degeneracies, i.e. the existence of various combinations of parameter values
that yield equally good results when reproducing a set of observed datapoints. Here we are
focusing on the properties of the AGN emission, and we will hence not face this issue with respect
to the stellar component (both stars and dust heated by star formation). We will exploit the fact
that the MIR domain is strongly dominated by AGN in our sample objects. With respect to this 
approach, the degeneracy problem translates into the fact that there are more AGN models, with
different parameters, that yield equally good --or acceptable in terms of $\chi^2$ values-- fits,
so that the properties of the dusty torus can not be unequivocally assessed.

To properly deal with this issue, we keep trace of the exploration of the model's parameter space,
and we analyse the 30 best model fits for each object of the sample.
For most of the objects, even when only the first two solutions are considered the discrepancies,
measured as $\Delta param/param(0)$ where $param(0)$ is the value of a given parameter
corresponding to the best fit model, are larger than 50\%. In order to understand this behaviour
one has to take into account the influence of each of the model parameters to the global model
SEDs as well as the fact that not all parameters are independent. In fact, most of them
are correlated in some way. For instance, the LoS and covering factor are tightly related:
if the covering factor is small, LoS can vary in a larger interval; however as the covering factor
increases, the LoS is restricted in those angles that permit direct view of the central source,
especially in the case of high optical depths.

\begin{figure}
\centerline{
\psfig{file=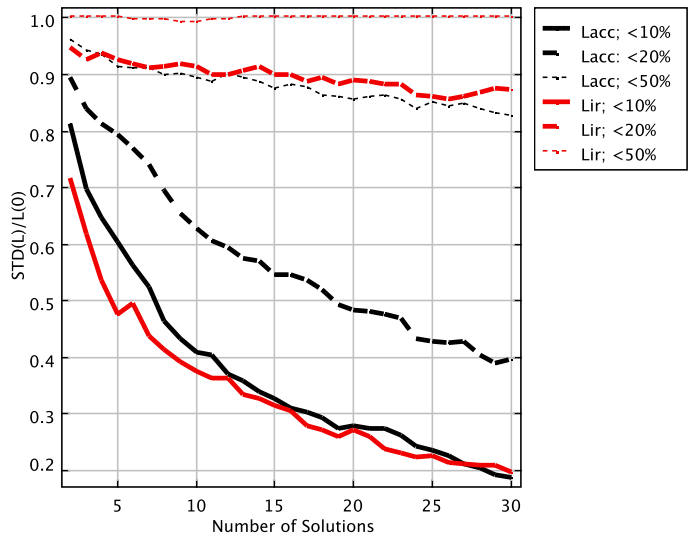,width=10cm}}
\caption{Percentage of objects with $\sigma$(L)/L(0) below given thresholds, 
as function of the number of solutions. Black lines: $\rm L_{acc}$;
red lines: $\rm L_{IR}$. The full, dashed and dotted lines represent the following thresholds: 
$\sigma$(L)/L(0) of $\le$ 10\%, 20\%, and 50\%, respectively.}
\label{fig:Lstats}
\end{figure}

In fact, the parameters that are best constrained are $\rm L_{acc}$ and therefore $\rm R_{in}$
(from equation \ref{eqn:rin}) and $\rm L_{IR}$, despite the lack of points longward $\rm \lambda$= 24
\mum. Fig. \ref{fig:Lstats} shows how the fraction of objects with standard deviation of
the luminosity over the best fit luminosity, $\sigma$(L)/L(0), of $\le$ 10\% (plain lines), 
$\le$ 20\% (dashed lines), and $\le$ 50\% (dotted lines), varies as a function of the
first $n$ solutions ($n$ ranges from 2 to 30).
$\rm L_{acc}$ is shown in black while $\rm L_{IR}$ is represented in red.

\section{DISCUSSION}
\label{discuss}

In this work we compare observed and model SEDs of SDSS quasars with SWIRE
counterparts aiming at constraining the model parameters and at quantifying the IR
properties of bright quasars. The need for such an approach, though evident, was also stressed by
\cite{gallagher07}, who studied a sample of 234 SDSS quasars, most of which also belong to
our sample, trying to quantify the effects of the luminosity on the shape of their 
SEDs in the MIR. Their conclusions were solely based on the observed SEDs and claimed
that comparison with models would constrain physical parameters, many of which are dealt 
with in the present study. 

We use a torus model with smooth dust distribution, originally presented in \cite{fritz06},
even though there are conflicting suggestions in the literature as to whether tori could be
smooth or clumpy. Smooth models have been tested on a variety of
objects (e.g. \citealt{granato94}; \citealt{fritz06}) yielding very good results.
There is, however, evidence that clumpy tori might be more realistic,
(e.g. \citealt{risaliti02}), nevertheless, the few models in the literature
were tested only on average SEDs \citep{nenkova02} or
on individual objects \citep{hoenig06}. Furthermore, \cite{nenkova02}
explicitly addressed the issue of suppression of the 9.7 \mums silicate feature,
predicted in emission from smooth models, which was never till then
supported observationally in type-1 objects, but the picture has changed since
with Spitzer IRS observations of this feature in emission (\citealt{siebenmorgen05}; 
\citealt{sturm05}; \citealt{hao05}; \citealt{buchanan06}; \citealt{shi06}).

\begin{figure}
\centerline{
\psfig{file=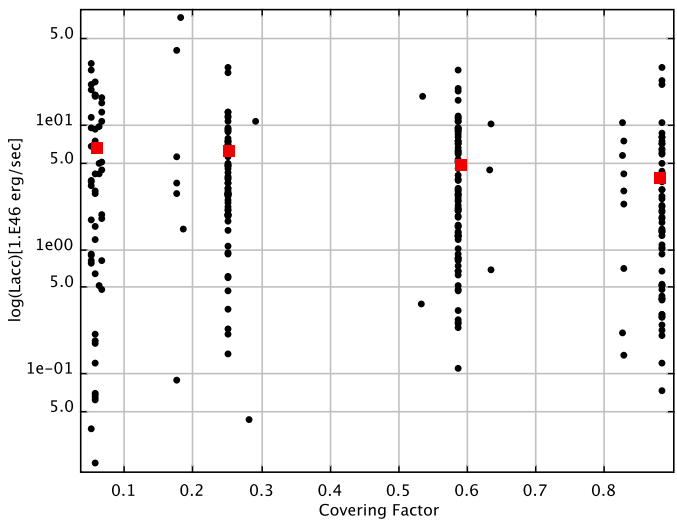,width=10cm}}
\caption{Accretion luminosity, $\rm L_{acc}$, versus CF: the mean $\rm L_{acc}$
(red squares) decreases in the bins with increasing CF but with large dispersions on the
values of $\rm L_{acc}$.}
\label{fig:LaccCF}
\end{figure}

The issue of the influence of the accretion luminosity on the dust coverage
is the source of an ongoing discussion in the literature.
According to the receding torus paradigm
\citep{lawrence91} the opening angle of the torus depends on the power
of the central sources, with more powerful quasars sweeping larger amounts of dust
leaving thus larger opening angles around them. 
In this approach, the Unified Scheme does not rely uniquely on the orientation of
the dusty torus but on the dependence of the geometrical thickness and the optical depths
on the central source. Evidence for receding tori is often found in studies
of radio-loud AGN (e.g. Grimes, Rawlings \& Willott 2003, 2005). Recently,
\cite{maiolino07} presented results of a study of high luminosity quasars carried out
with Spitzer IRS pointing towards decreasing covering factors with increasing
luminosity.
In Section \ref{sec:lir} we already suggested that the influence of $\rm L_{acc}$
on the torus geometry is demonstrated by the larger values of $\rm L_{acc}$ generaly occuring in
objects with $\gamma=-6.0$. In fact, $\rm L_{acc}$ as derived from the best fit models,
also shows a slight dependence on CF with the average value of $\rm L_{acc}$ in bins of
covering factor (red squares) decreases with increasing covering factor, as seen in Fig.
\ref{fig:LaccCF}. The dispersion on $\rm L_{acc}$ in each bin, however, is so large
that the study is inconclusive.

The study of the sources with 70 \mums detections, representing 25\% of the sample, 
suggested the presence of a strong starburst component in many of the cases.
According to the best fit models and due to the flux limits of the sample, only 
objects with Arp220-like components were observed in 70 \mum. 
One could argue that Arp220 is a heavily extinguished starburst, prototype 
ultraluminous IR galaxy in the local Universe and therefore unlikely to be found in
high redshift bright quasars. In a simpler approach, a black body of $\sim$30K could have
been used to account for the 70 micron detection but we used observed starburst templates
instead, in the line of \cite{fritz06}. However, we did exclude from the model grid models with 
R$_{\rm out}/$R$_{\rm in}$=300, originally included in \cite{fritz06}. These models imply larger
dust distributions and lower dust temperatures and could have been used to reproduce the
70 \mums detections. However, they would also
correspond to tori with physical sizes of several hundred parsecs, sometimes 
even kpc, where star formation is likely to occur. Furthermore, at those distances the dust 
temperature would drop to temperatures typical of the dust of starburst or the diffuse 
dust responsible for cirrus emission.

One of the main points of this work is the use of tori models with low equatorial
optical depths at 9.7\mum, $\tau_{9.7} < 1.0$. Our study concludes that the presence 
of low optical depth tori around active nuclei is a possibility, 
and would imply only a minimum of modifications in the current picture of the Unified 
Scheme, namely the possibility of seeing type-1 objects while dust intercepts the line
of sight. A strong implication of this assumption would be the presence of the silicate
feature at 9.7 \mums in emission even in type-2 AGN. Such a feature was indeed observed
recently in X-ray selected type-2 AGN, using the Spitzer IRS \citep{sturm06}.
The comparison between results yielded by all $\tau_{9.7}$ models and only high
$\tau_{9.7}$ ones suggests a flattening of the distribution
of the covering factor (see Fig. \ref{fig:histoCF}, increasing thus the apparent
estimated ratio between type-2 and type-1 objects, as dust enshrouded AGN
could still be seen as type-1s if seen through a low optical depth medium.
In order to really probe the Unified Scheme and test the validity of the models, on would need
to select a complete, volume limited sample and study {\it all} AGN inside this volume,
i.e. of both types 1 and 2. The comparison of the tori model parameters and the AGN properties
between the two types would prove or not the validity of the Unification Scheme, since
the distribution of the various quantities should be the same for both types. Furthermore,
the SED fitting could provide information about the orientation of the sources and finally the
number of obscured versus unobscured quasars. This is the subject of our forecoming work.

\subsection{On the limitations of the models and results}
\label{limits}

It has been argued on many occasions that SED fitting techniques are both
powerful {\it and} limited. Powerful because photometry is and will always be 
available for samples that are orders of magnitudes larger than spectroscopic ones;
limited because it greatly depends on the quality of the photometric data, the
model or synthetic SEDs used and the way of populating the multi-parameter space, 
the aliasing and degeneracy issues, and the priors one imposes.
In the majority of cases, the value of SED fitting of large datasets
is statistical, it helps define trends of the population under study 
as a whole, rather than the properties of individual objects, unless the sampling
of their observed SEDs and the nature of the objects themselves allow for it.

Quasars in general have unmistakable signatures both in the UV/optical (in general 
blue continua attributed to the accretion onto the central black hole) and in the IR 
(IR bump due to dust torus emission). Furthermore, the sample under study has
good wavelength coverage with SDSS, 2MASS and SWIRE data available, as
detailed in Sections \ref{data} and \ref{results}, helping pin down the wavelength
where the falling emission from the nucleus meets the rising emission from the dust
(around 1 micron).

Since we are working with a pre-constructed grid of models, all of the models 
input parameters take discrete values and, in some cases, the model parameter
spaces is sparsely populated. This is the case, for instance, of the assumed
$\rm R_{out}/R_{in}$ ratios, taking two discrete values of 30 or 100, or the 
angular variation of the dust density (factor $\gamma$ in equation 
\ref{eqn:density}), being either 0.0 or -6.0. Recalculating a model to
match each individual object would probably yield better results, however this
is computationally very demanding and sometimes even impossible, as some of
the model parameters combination (very high optical depth, $\rm \tau_{9.7}>6.0$
and steep density profiles) would lead to a non-convergence of the
torus model (see \citealt{fritz06}). This, of course, limits the accuracy of
the results and therefore the values of the physical parameters derived should
be seen as indicative. 

\noindent
\vspace{0.75cm} \par\noindent
{\bf ACKNOWLEDGMENTS} \par

\noindent This work is based on observations made with the {\it Spitzer Space Telescope},
which is operated by the Jet Propulsion Laboratory, California Institute of
Technology under NASA contract 1407.
Support for this work, part of the Spitzer Space Telescope Legacy Science
Program, was provided by NASA through an award issued by the Jet Propulsion
Laboratory, California Institute of Technology under NASA contract 1407.

Funding for the creation and distribution of the SDSS Archive has been provided
by the Alfred P. Sloan Foundation, the Participating Institutions, the National
Aeronautics and Space Administration, the National Science Foundation, the U.S.
Department of Energy, the Japanese Monbukagakusho, and the Max Planck Society.
The SDSS Web site is http://www.sdss.org/.

This publication makes use of data products from the Two Micron All Sky Survey, 
which is a joint project of the University of Massachusetts and the Infrared 
Processing and Analysis Center/California Institute of Technology, funded by the 
National Aeronautics and Space Administration and the National Science Foundation.

This work was supported in part by the Spanish Ministerio de
Educaci\'on y Ciencia (Grants ESP2004-06870-C02-01 and ESP2007-65812-C02-02)

This work makes extensive use of TOPCAT (http://www.starlink.ac.uk/topcat/)
a Virtual Observatory tool developed by M. Taylor.

\end{document}